\documentclass[preprint2]{aastex6}
 \usepackage{color}

\newcommand\erf{\rm erf}
\fullcollaborationName{The Friends of AASTeX Collaboration}

\begin{document}
\title{An analytical model of radial dust trapping in protoplanetary disks}
\author{Anibal Sierra \altaffilmark{1}, Susana Lizano \altaffilmark{1}, Enrique Mac\'ias \altaffilmark{2}, Carlos Carrasco-Gonz\'alez \altaffilmark{1}, Mayra Osorio \altaffilmark{3}, Mario Flock \altaffilmark{4}}
\altaffiltext{1}{Instituto de Radioastronom\'ia y Astrof\'isica, UNAM, Apartado Postal 3-72, 58089 Morelia Michoac\'an, M\'exico}
\altaffiltext{2}{Department of Astronomy, Boston University, 725 Commonwealth Avenue, Boston, MA 02215, USA}
\altaffiltext{3}{Instituto de Astrof\'isica de Andaluc\'ia (CSIC) Glorieta de la Astronom\'ia s/n E-18008 Granada, Spain}
\altaffiltext{4}{Max Planck Institute f\H{u}r Astronomy (MPIA), K\H{o}nigsthul 17, 69117 Heidelberg, Germany}

\begin{abstract}
We study dust concentration in axisymmetric gas rings in protoplanetary disks. Given the gas surface density, 
we derived an analytical total dust surface density by taking into account the differential concentration of all the grain sizes.  This model 
allows us to predict  the local dust-to-gas mass ratio and the slope of the particle size distribution, as a function of radius.
We test this analytical model comparing it with a 3D magneto-hydrodynamical simulation of dust evolution in an accretion disk. 
The model is also applied to the disk around HD 169142. By fitting the disk continuum observations simultaneously at $\lambda = 0.87$, 1.3, 3.0 mm, we obtain a global dust-to-gas mass ratio $\epsilon_{\rm global} = 1.05 \times 10^{-2}$ and a viscosity coefficient $\alpha = 1.35 \times 10^{-2}$. This model can be easily implemented in numerical simulations of accretion disks.

\end{abstract}


\section{Introduction}
In the last few years, high quality (sub)mm observations of disks at high angular resolution have shown that a significant fraction of them hosts one or more concentric rings and gaps (e.g. \cite{ALMA_2015}, \cite{Andrews_2018}). These radial structures have been seen in both, the gas and the dust (e.g. \cite{Isella_2016}). Although their origin is still under debate (see e.g. \cite{Carrasco_2016} and references therein), these structures must have a strong impact on the evolution of the dust and gas. Ultimately, understanding this evolution is fundamental to figure out how the formation of planetary systems takes place.

Dust grains migrate radially in protoplanetary disks due to their interaction with the gas molecules: the radial pressure gradient felt by the gas is responsible for the shear in the angular velocity between the dust grains and the gas molecules, which causes an angular momentum interchange between both components via a drag force. 
The dust radial velocity is proportional to the radial pressure gradient (e.g. \cite{Whipple_1972}, \cite{Weidenschilling_1977}).
For a disk with decreasing density and temperature radial profiles, the pressure gradient  points radially inward and causes a radial migration of the dust grains toward the central star. In addition, the magnitude of the velocity depends on the size of the dust grains (e.g., \citep{Takeuchi_2002}). Small dust grains ($\sim \mu$m) are well coupled with the gas, and large grains with sizes less than $\sim 1$ meter feel a strong drag force and are expected to accrete to the star in a timescale much less than the lifetime of the gaseous disk, this problem is known as the radial drift barrier, first discussed by \cite{Weidenschilling_1977}.

If the gas pressure has local maxima, as in the case of vortices (e.g., \cite{Barge_1995}) or rings (e.g., \cite{Pinilla_2012}), the dust grains could be trapped, avoiding or retarding their accretion toward the star, promoting the appropriate conditions to build planetesimals.

\cite{Pohl_2017} recently studied the disk around HD 169142 where two main gaps, probably created by Jupiter-mass planets, perturb the local gas  and trap dust grains in the outer zone of the gaps, where the gas pressure has a maximum. Using their dust evolution model that includes migration and grain growth, they can explain the $1.3$ mm continuum emission of the disk observed by \cite{Fedele_2017}, showing that the millimeter grain sizes have had an important evolution (migration + growth) within the disk. Pohl et al. also modeled the population of small dust grains in order to explain the dust scattering at near-IR wavelengths. This small dust population is well mixed with the gas. 
  
In this paper we study from an analytic point of view how the dust grains could be trapped in axisymmetric rings where the gas pressure has local maxima, and apply this model to the disk around HD 169142. In this approach, the dust grains  radially migrate toward the pressure maxima and reach an equilibrium with the turbulent mixing which depends on the grain size, but they do not grow. We do not address what creates the gas gaps (which in turns creates the gas pressure maxima), but we are only interested in the redistribution of dust grains given the gas surface properties, i.e., the dust dynamics is only influenced by the drag force and turbulent mixing. Possible gravitational interactions with planets within the gaps are also ignored.

The structure of this paper is as follows: Section \ref{SEC:Analytical_model} describes the analytical model of dust concentration within the  gas pressure maxima based on the dust dynamics within protoplanetary disks. Section \ref{SEC:SIMULATION} tests the analytical dust model by comparing it with recent dust and gas disk simulations. This model is then used to describe the dust emission of the disk around the HD 169142 star (Section \ref{SEC:Model_HD169142}). The observational properties of the disk are summarized in Subsection \ref{SUBSEC:Observations}. In Subsection \ref{SUBSEC:Gas_surface_density} we derive a model for the HD 169142 gas surface density, which is used together to the radiative transfer solution described in Subsection \ref{SUBSEC:Radiative_transfer} in order to obtain the dust surface density from a best fit model  by comparing with the disk emission at multi-wavelength millimeter observations (Subsection \ref{SUBSEC:Model_grid}). 
In Section \ref{SEC:amax_p_degeneration} we discuss the degeneration between the dust properties that can produce the same observed value of the spectral index. Conclusions are presented in Section \ref{SEC:Conclusions}.

\section{Analytical model}
\label{SEC:Analytical_model}
Dust grains can be described as a pressure-less fluid, so their evolution in a thin disk can be followed by the advection-diffusion equation
\begin{equation}
\frac{\partial \Sigma_{\rm d}}{\partial t} = \nabla \cdot \left( \Sigma_{\rm d} \vec{v} \right) - \nabla \cdot \left[ D \Sigma_{\rm g} \nabla \left( \frac{\Sigma_{\rm d}}{\Sigma_{\rm g}} \right) \right],
\label{Eq:Difussion_eq}
\end{equation}
where $D$  $\{ \rm cm^2 s^{-1}\}$ is the dust diffusion coefficient, $\Sigma_{\rm d}$ and $\Sigma_{\rm g}$ are the dust and gas surface densities, respectively, and $\vec{v} = (v_{\varpi}, v_{\phi})$ is the dust velocity in cylindrical coordinates $(\varpi, \phi)$. A steady state solution can be reached if the flux of dust grains toward pressure maxima is balanced by diffusion, in contrast with disks 
where the pressure monotonically decreases with the radius and where the dust grains always migrate radially inward. In an axisymmetric disk with local pressure maxima, the steady state solution of Equation (\ref{Eq:Difussion_eq}) is given by
\begin{equation}
0 =  \Sigma_{\rm d} v_{\varpi} - D \Sigma_{\rm g} \frac{d}{d \varpi}\left( \frac{\Sigma_{\rm d}}{\Sigma_{\rm g}} \right).
\label{Eq:Equilibrium}
\end{equation}
This equation assumes that the dust follows the gas, i.e., that there is not back reaction of the dust on the gas. This effect can be neglected if  the dust-to-gas mass ratio is less than 1 (e.g., \cite{Taki_2016}).

In the Epstein regime (where the radii of the dust grains are smaller than 4/9 of the mean free path of the gas), the radial dust velocity is given by (e.g., \cite{Windmark_2012})
\begin{equation}
v_{\varpi} = \left(\frac{St}{1+St^2} \right)\frac{1}{\Omega \Sigma_{\rm g}} \frac{d P}{d \varpi},
\label{EQ:dust_vel}
\end{equation}
where 
\begin{equation}
St = \frac{\pi \rho_m a}{2\Sigma_{\rm g}},
\label{EQ:Stokes_number}
\end{equation}
is the Stokes number of a dust grain with radius $a$ and material density $\rho_m$, and $P = \Sigma_{\rm g} c_s^2$ is the gas pressure, where $c_s$ is the sound speed.
Assuming that the dust diffusion coefficient is $D = D_{\rm g}/(1+St^2)$ \citep{Youding_2007}, where $D_{\rm g} = \alpha c_s^2/\Omega$ is the gas diffusion coefficient, $\Omega$ the angular speed and $\alpha$ the viscosity coefficient, then, the Equation (\ref{Eq:Equilibrium}) can be written as
\begin{equation}
0 = \frac{d}{d \varpi} \left( \frac{\Sigma_{\rm d}}{\Sigma_{\rm g}} \right) - \frac{St}{\alpha c_s^2 \Sigma_{\rm g}} \frac{d P}{d \varpi} \left( \frac{\Sigma_{\rm d}}{\Sigma_{\rm g}} \right). 
\label{Dtemp}
\end{equation}

Neglecting the thermal gradients with respect to the density gradients,
\footnote{For typical sound speed and gas surface density gradients, $ {d \ln c_2^2  / d \varpi} \sim -1/2$ and
${d\ln  \Sigma_{\rm g} /  \varpi } \sim -1$, there would be an correction term of 3/2 in front of the second term in eq. (\ref{Dtemp}). However, for a disk with gaps, the difference tends to be larger ${|d\ln  \Sigma_{\rm g} /  d\varpi |}  \gg {|d \ln c_2^2  / d \varpi |} $. 
} the solution of the above equation is
\begin{equation}
\frac{\Sigma_{\rm d}(\varpi)}{\Sigma_{\rm g}(\varpi)} = \frac{\Sigma_{\rm d}(\varpi_0)}{\Sigma_{\rm g}(\varpi_0)} \exp \left[ \int_{\varpi_0}^{\varpi} \frac{St}{\alpha} \frac{d \ln( \Sigma_{\rm g})}{d \varpi} d\varpi \right],
\label{Eq:Eq_2Solve}
\end{equation}
where $\Sigma_{\rm d}(\varpi_0), \Sigma_{\rm g}(\varpi_0)$ are the dust and gas densities at a reference radius $\varpi_0$. 
Finally, the dust surface density can be written as 
\begin{equation}
\Sigma_{\rm d}(\varpi,a) = \epsilon_0 \Sigma_{\rm g}(\varpi) \exp \left[ - ka \right],
\label{EQ:Dust_model}
\end{equation}
where
\begin{equation}
k = \frac{\pi \rho_m}{2\alpha} \left( \frac{1}{\Sigma_{\rm g}(\varpi)} - \frac{1}{\Sigma_{\rm g}(\varpi_0)} \right) ,
\label{EQ:k}
\end{equation}
and $\epsilon_0$ is the dust-to-gas mass ratio at the reference radius $\varpi_0$. 
Figure (\ref{FIG:Dust_densities}) shows the normalized dust surface density for grains of different sizes $a [\rm cm]= 10^{-4}, 10^{-3}, 10^{-2}, 10^{-1}, 10^{0}$, $\alpha = 10^{-2}$, and a gas surface density with two local maxima at $\varpi = 24$ AU and $60$ AU (black dashed line).
This profile corresponds to the gas surface density profile of HD 169142 discussed in subsection \ref{SUBSEC:Observations} below. Note that the largest dust grains are more concentrated around the first maxima. As the radii of the grains decrease, the dust particles tend to be well mixed with the gas. The grains with a size smaller than $10^{-3}$ cm (yellow dashed line) trace the gas surface density.  

Equation ({\ref{EQ:Dust_model}) can also be written as  
\begin{equation}
\epsilon(\varpi,a)/\epsilon_0  \propto \exp[-1/\Sigma_{\rm g}(\varpi)],
\end{equation}
where $\epsilon(\varpi, a)$ is the local dust-to-gas mass ratio.  
Therefore the local dust to gas mass ratio follows the gas extrema (maxima or minima).
 Equation ({\ref{EQ:Dust_model}) also predicts that the gaps in the dust surface density are deeper than in the gas surface density. Note that, because the dust mass is conserved, the maxima are enhanced due to the redistribution of the dust grains. 

\begin{figure}
\includegraphics[scale=0.5]{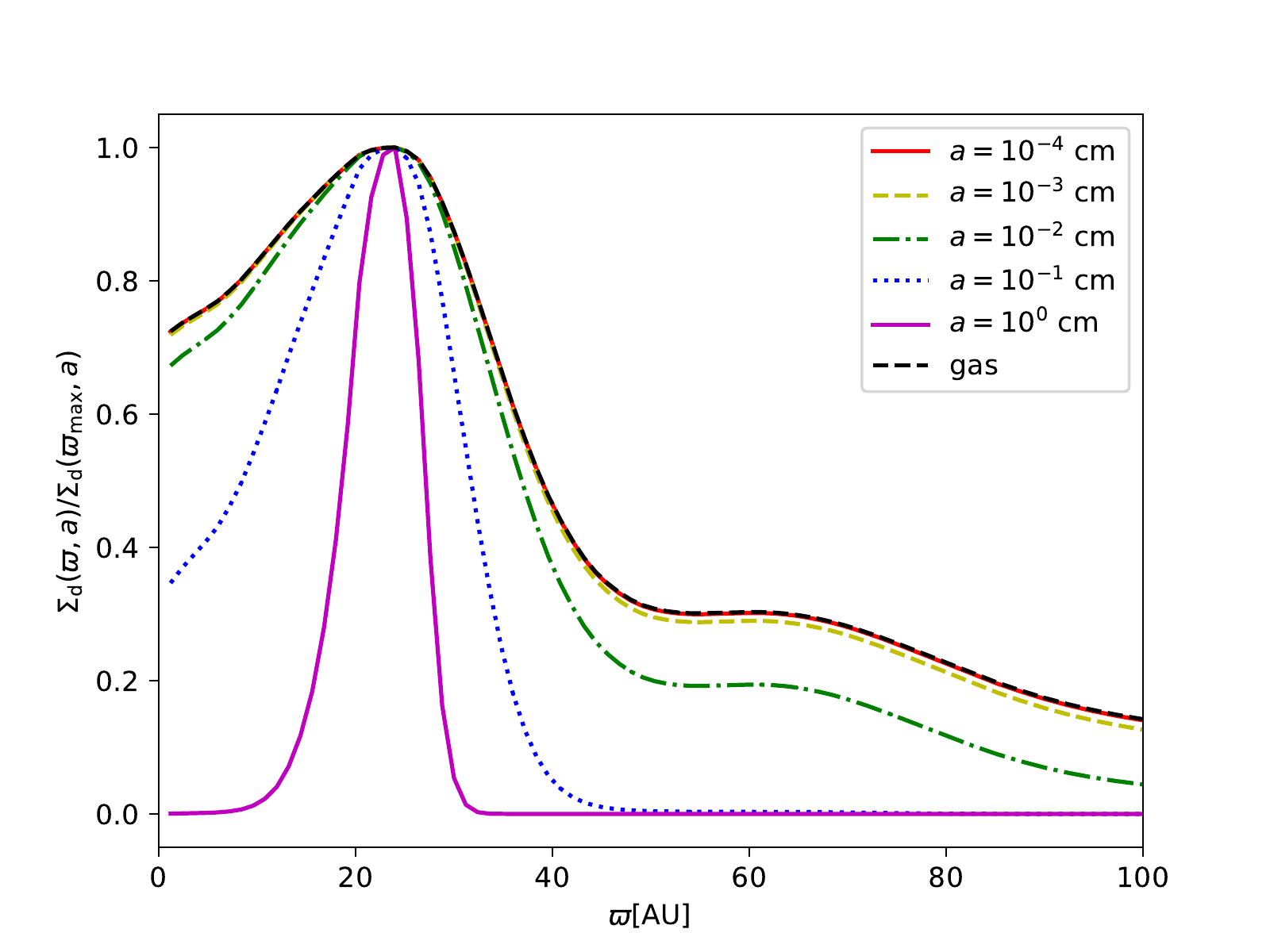}
\caption{Normalized dust surface density for grains of different sizes.}
\label{FIG:Dust_densities}
\end{figure}

The total dust surface density is given by the sum of the dust densities for each grain size. 
For a dust size distribution where $n(a)da \propto a^{-p}da$  is the number of dust grains per unit volume with a radius between 
$a$ and $a+da$, with  minimum and maximum grain sizes,  $a_{\rm min}$ and $a_{\rm max}$, respectively,  the total dust surface
density is
\begin{equation}
\Sigma_{\rm d}(\varpi) = \frac{\int_{a_{\rm min}}^{a_{\rm max}} \Sigma_{\rm d}(\varpi, a) a^{3} n(a) da }{\int_{a_{\rm min}}^{a_{\rm max}} a^3 n(a)da}.
\label{EQ:Total_dust_density}
\end{equation}

In this equation, the factor $a^3n(a) da$  weights the amount of mass associated to each dust grain size, which is assumed to be constant because the dust grains are only redistributed in the disk without coagulation or fragmentation. In addition, the redistribution of the dust grains is taken into account by the factor $\Sigma_{\rm d}(\varpi,a)$,
which includes the dust size differential migration,  see Figure (\ref{FIG:Dust_densities}).
\footnote{The local changes in the particle size distribution are discussed in Section (\ref{SEC:amax_p_degeneration}).}
Thus, the integral can be computed using the global values of the disk 
($a_{\rm max}, p$). 

In particular, if the original value of the particle size distribution is $p=3.5$ and $a_{\rm max} \gg a_{\rm min}$, the integral is
\begin{equation}
\Sigma_{\rm d}(\varpi) = \frac{\epsilon_0 \Sigma_{\mathrm{g}}}{2\sqrt{ka_{\mathrm{max}}}} \erf(\sqrt{ka_{\rm{max}}}),
\label{EQ:Sigma_dust_nonorm}
\end{equation}
where $\erf$ is the error function. 
Furthermore, $\epsilon_0$ is constrained by the total dust mass in the disk. If $\epsilon_{\rm global}$ is the global dust-to-gas mass ratio (typically $\sim 1/100$), then
\begin{equation}
\epsilon_{\rm global} = \frac{\int_A \Sigma_{\rm d}(\varpi) dA}{\int_A \Sigma_{\rm g}(\varpi)dA} = \frac{\int_A \Sigma_{\rm d}(\varpi) dA}{M_{\rm g}},
\end{equation}
where $M_{\rm g}$ is the total gas mass. Therefore, using  Equation (\ref{EQ:Sigma_dust_nonorm}) one obtains
\begin{equation}
\epsilon_0 = \epsilon_{\rm global} M_{\rm g} \left[ \pi \int_0 ^{R} \frac{\Sigma_{\rm g}(\varpi) \erf(\sqrt{ka_{\rm max}})}{\sqrt{ka_{\rm max}}} \varpi d\varpi \right]^{-1},
\label{EQ:Dust_constain}
\end{equation}
where $R$ is the disk radius. Finally, replacing Equation (\ref{EQ:Dust_constain}) in  Equation (\ref{EQ:Sigma_dust_nonorm}) one obtains the total dust surface density as a function of the gas surface density, the maximum grain size and the global dust-to-gas mass ratio as
\begin{equation}
\Sigma_{\rm d}(\varpi) = \frac{\epsilon_{\rm global} M_{\rm g}}{2\pi} \left[ \frac{\Sigma_{\rm g} \erf (\sqrt{ka_{\rm max}}) / \sqrt{ka_{\rm max}} }{\int_0^R \Sigma_{\rm g} \erf (\sqrt{ka_{\rm max}}) / \sqrt{ka_{\rm max}}\varpi d\varpi }   \right].
\label{EQ:Total_dust_surface_density}
\end{equation}
This equation gives the dust surface density when the gas surface density in axisymmetric rings is known. Since the gas surface density evolves in a diffusion timescale which is much longer than the advection timescale of the dust (See Appendix \ref{APP:Timescales}), the dust will always concentrate following the gas maxima.

Recently, \cite{Dullemond_2018arXiv} proposed a dust model that assumes that the gas pressure maxima are given by gaussian functions. To obtain the dust surface density of grains with a single size, they assume locally a maximum gas surface density given by Toomre stability criterion.
Instead, our model uses a gas profile that can be obtained from observations or simulations. Given this profile, it predicts the total 
dust surface density taking into account the differential concentration of all grain sizes. For non-axisymmetric disks, \cite{Sierra_2017} modelled the total dust concentration in disk vortices taking into account the distribution of dust grains sizes.

\section{Application of the model to a numerical simulation}
\label{SEC:SIMULATION}
\cite{Flock_2015}, \cite{Ruge_2016} performed non-ideal 3D magneto-hydrodynamical simulations of a protoplanetary disk where they
followed the dynamics of dust particles of different sizes. As expected, they found that the largest grains tend to be more concentrated around the pressure maxima. In this section we compare the dust surface density from the simulation for each grain size with that predicted by our analytic dust model (Equation \ref{EQ:Dust_model}).  Then, the gas surface density that appears in the dust model equation is given by the azimuthally averaged gas density profile from the simulation.

We choose a snapshot of the simulation at a time of 400 inner orbits of the disk. At this time the gas surface density has developed a ring structure without vortices\footnote{The azimuthal fluctuations of the gas surface density in the ring with respect to the azimuthal average are smaller than 6\%.}, 

and the dust grains have had enough time to concentrate in the ring following the drag force of the gas. The gas density profile has a local maximum (ring)  centered at $65$ AU. 
We  consider the gas density profile between $55$ to $85$ AU,  where the radial drift has stopped and the dust has achieved a steady state.

The concentration of the dust surface density depends on the assumed value of the viscosity coefficient $\alpha$ (Equation \ref{EQ:k}) which is found from the best fit model  by minimizing the function
\begin{equation}
\sigma^2 = \frac{1}{N_a} \frac{1}{N_p} \sum_a \sum_{\varpi}
\left[ \Sigma_{\rm d,norm}^{\rm sim}(\varpi,a) - \Sigma_{\rm d,norm}^{\rm mod} (\varpi,a) \right]^2;
\end{equation}
where $N_a$ is the number of dust particle sizes, and $N_p$ is the number of radial points sampled, $\Sigma_{\rm d,norm}^{\rm mod}$ is the normalized dust surface density profile of the model, and $\Sigma_{\rm d,norm}^{\rm sim}$ is the normalized azimuthally-averaged dust density profile of the simulation.

The sum over $\varpi$ compares the radial profiles and the sum over $a$ take into account all the grain sizes, which vary from 50 $\mu$m to $1.0$ cm in 10 logarithmically equally spaced bins. Figure (\ref{FIG:Best_alpha}) shows $\sigma^2$ as a function of $\alpha$. The minimum is obtained for $\alpha_b = 1.3 \times 10^{-3}$, which is of the same order of magnitude of the averaged value, $3\times 10^{-3}$, found in the simulation.
The value of dust-to-gas mass ratio $\epsilon_0$ is not fitted because the dust particles included in the simulation are only a representative sample of the total dust mass; that is the reason why we compare the normalized dust surface densities.

\begin{figure}
\includegraphics[scale=0.5]{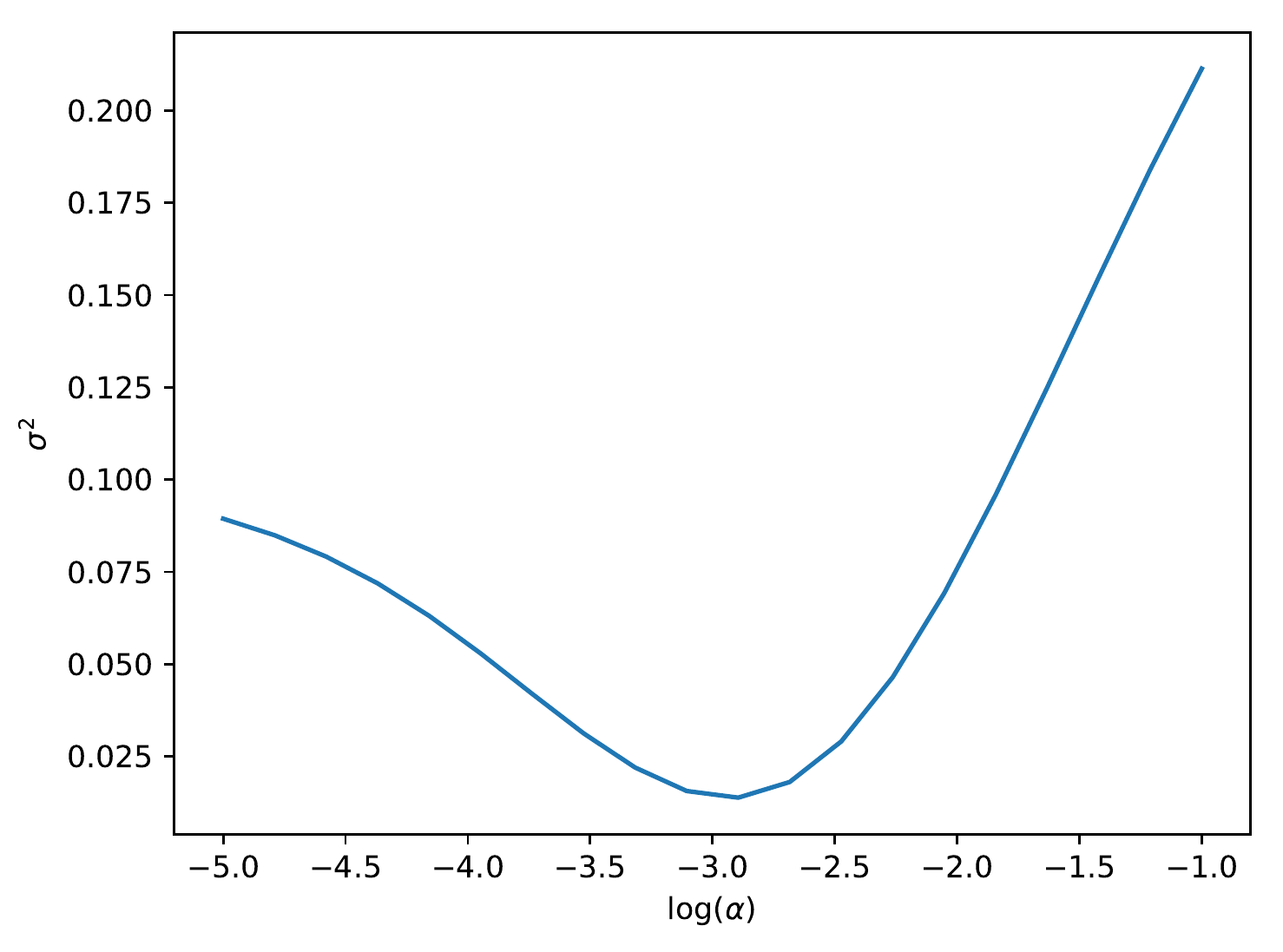}
\caption{Error estimation between the analytic model and simulations as a function of $\alpha$. }
\label{FIG:Best_alpha}
\end{figure}

Figure (\ref{FIG:Best_sim}) shows the normalized surface densities from the model (solid lines) and the simulation (dotted lines) using the viscosity coefficient $ \alpha_b$ and for three representative dust grain sizes: $a = 0.05$ cm (red line), $0.09$ cm (green line), and $1.0$ cm (blue line). In all cases the width of the model profiles  match the simulation
profiles.

\begin{figure}
\includegraphics[scale=0.5]{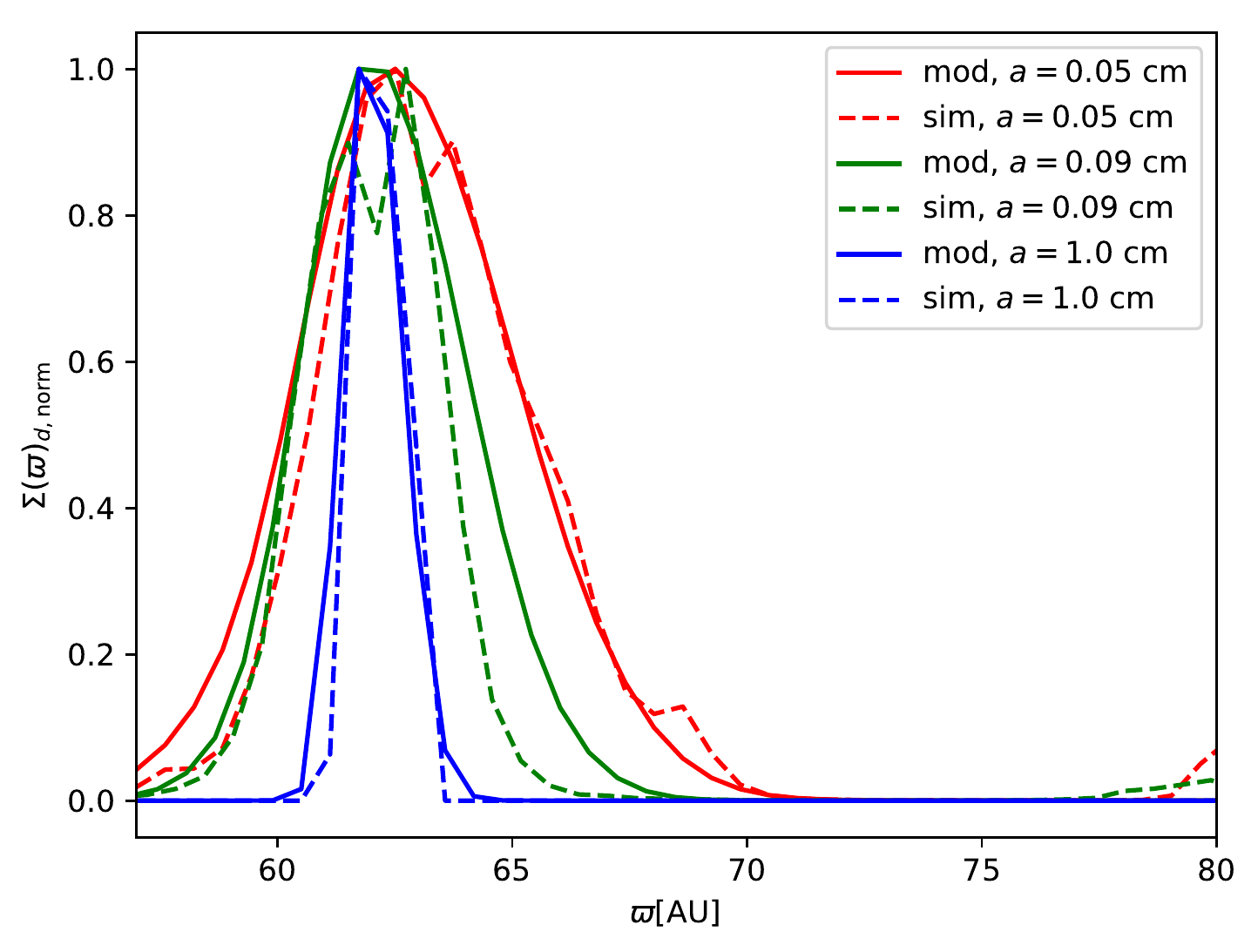}
\caption{Normalized surface density from the model (solid lines) and the simulation (dotted lines) for grains with sizes $a = 0.05$ cm (red line), $0.09$ cm (green line) and $1.0$ cm (blue line).}
\label{FIG:Best_sim}
\end{figure}

\section{Application of the model to HD 169142}
\label{SEC:Model_HD169142}
In this section we applied the analytical model (Equation \ref{EQ:Total_dust_surface_density}) that gives
the dust surface density $\Sigma_d(\varpi)$ to simultaneously explain the millimeter dust emission of the disk around HD 169142 at different wavelengths $\lambda = 0.87, 1.3, 3.0$ mm, using the observed gas properties.

Subsection \ref{SUBSEC:Observations} summarizes the observed properties of the central star and the disk in this source. In subsection \ref{SUBSEC:Gas_surface_density} we derive the disk gas surface density and excitation temperature 
from the $^{12}$CO and $^{13}$CO maps.  These disk gas properties are used in the analytical model to predict the
dust surface density, in order to  determine the best fit values for the global dust-to-gas mass ratio ($\epsilon_{\rm global}$) and the viscosity coefficient ($\alpha$). We choose the parameters that minimize the reduced chi-squared of the dust continuum emission of the observed dust maps and the emission of a grid of  models. Subsection \ref{SUBSEC:Radiative_transfer} describes how to compute the emergent specific intensity of the models and subsection \ref{SUBSEC:Model_grid} presents the results for the best values of $\epsilon_{\rm global}$ and $\alpha$.
\subsection{HD 169142}
\label{SUBSEC:Observations}
HD 169142 is a Herbig Ae star \citep{The_1994} in the Sagittarius constellation with an age of $10$ Myr \citep{Pohl_2017}, a mass of $M_* =  1.65 M_{\odot}$ \citep{Blondel_2006} with an effective temperature $T_* = 8100$ K, a star radius of $R_* = 2.2 R_{\odot}$ \citep{Osorio_2014}, and at a distance of $d = 117 \pm 4$ pc \citep{Gaia_2016}.
The disk around HD 169142 is seen almost face-on, with an inclination angle of $\sim$ 14$\deg$ ($\cos i = 0.974$), and with its major axis along a position angle P.A. $= 5 \deg$ \citep{Raman_2006} on the plane of the sky. 

IR-polarized scattered-light images \citep{Quanz_2013} revealed that the disk has a central cavity, surrounded by a bright rim of radius $\sim 0.21''$ ($\sim$ 25 AU at 117 pc), and an annular gap ranging from $\sim0.28''$-$0.48''$ ($\sim$33-56 AU) in radius. An unresolved IR source was detected inside the central cavity (\citealt{Biller_2014}, \citealt{Reggiani_2014}) at a radius of $\sim 0.16''$ (19 AU) and was interpreted as a substellar or planetary companion candidate. The dust thermal emission of the disk was first imaged with the VLA at 7 mm (\citealt{Osorio_2014}, \citealt{Macias_2017}). These 7 mm observations confirmed the IR results and revealed that the ring of radius 25 AU is indeed narrow and azimuthally asymmetric, with a bright knot at PA= $-40 \deg$. These 7 mm images also suggest the presence of a possible new gap at radius $\sim$ 85 AU, located very close to the CO snowline, as imaged from DCO$^+$ ALMA data \citep{Macias_2017}. Finally, the VLA observations revealed a compact source of ionized material near the center of the cavity that could be tracing a weak radio jet, a photoionized inhomogeneous region of the inner disk, or an independent orbiting object. 

The HD 169142 disk has been observed with ALMA at $\lambda = 0.87$ mm, $1.3$ mm, and $3$ mm (archival data from program 2012.1.00799.S, \citealt{Fedele_2017}, and Macias et al. in prep. respectively). The submillimeter and millimeter continuum ALMA maps of the dust nicely confirmed the ringed structure of the disk initially revealed by the infrared polarized images and the 7 mm observations. The line emission of the gas has been observed with ALMA using $^{12}$CO, $^{13}$CO, C$^{18}$O $J: 2 \rightarrow 1$ rotational transitions by \cite{Fedele_2017}. 

In all the cases, the contrast between the rings and the gaps  in the line emission maps is weaker than in the continuum maps. For the $^{12}$CO this could be a consequence of the line emission being optically thick.  However, the $^{13}$CO emission, that is usually optically thin, also shows a smaller contrast compared to the dust emission; this behavior is similar to the dust model described in Section \ref{SEC:Analytical_model}.
Also, the line emission extends to a radius of $\sim 240$ AU, significantly larger than the radius inferred from the dust continuum emission, 
which only extends to $\sim 100$ AU.  This difference could be explained by the radial dust migration (e.g., \cite{Brauer_2008}).

Figure (\ref{FIG:HD169142_profiles}) shows the normalized azimuthally-averaged specific intensity of the dust continuum emission at $\lambda = 870 \ \mu$m (red solid line), $1.3$ mm (yellow dash-dotted line), and $3$ mm (green dashed line)\footnote{Table \ref{TAB:Obs_params} shows the parameters of the images at each wavelength.}. It also shows the normalized  azimuthally-averaged profiles the $^{13}$CO $J:2\rightarrow 1$ (cyan solid line) and the $^{12}$CO $J:2\rightarrow 1$ (blue dash-dotted line) line emission. All the dust maps were convolved to the same circular beam of 0.20 arcsec, while the gas maps have a circular beam of 0.16 arcsec. These beams are equivalent to $23.4$ and $18.7$ AU at the assumed distance of 117 pc respectively.  Note that, since the dust analytical model depends on the gas surface density (Equation \ref{EQ:Total_dust_surface_density}), the model can only resolve dust structures with the angular resolution of the gas maps; this is the reason why we choose a higher angular resolution in the gas; however, prior to compare the dust properties from the observations and the analytic model, we convolve the analytic dust model such that the final beam coincide with the observations.

\begin{figure}[!t]
\centering
\includegraphics[scale=0.5]{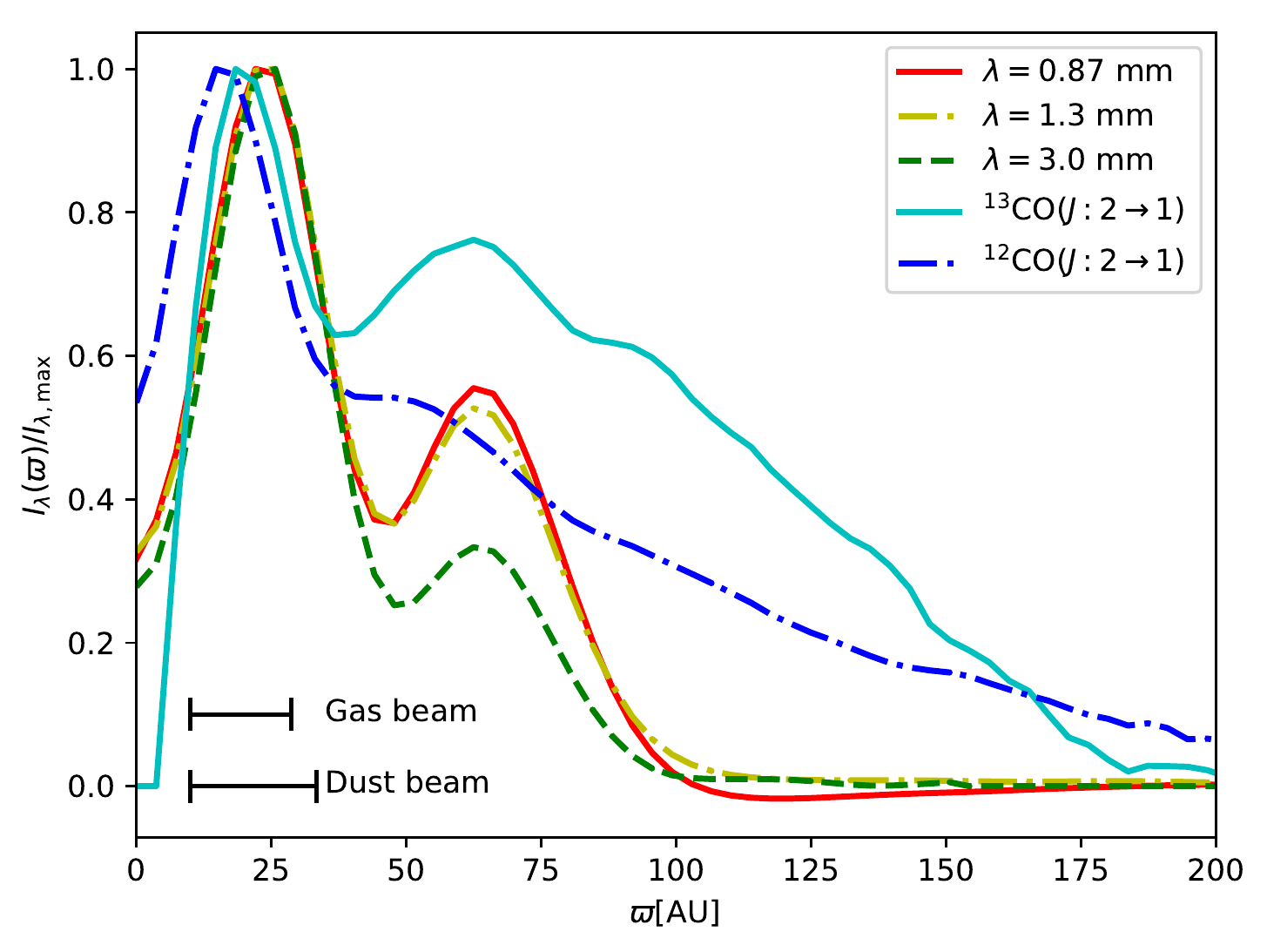}
\caption{Normalized HD 169142 specific intensities as a function of disk radius. The continuum dust emission at $\lambda = 0.87, 1.3,$ and $3.0$ mm are represented by the red solid line, yellow dash-dotted line, and green dashed line respectively. The line gas emission of the $^{13}$CO and $^{12}$ CO molecules are the cyan solid line and the blue dash-dotted line respectively. The dust emission maps have been convolved to a beam f 0.2 arcsec, while the gas maps were convolved to a beam of 0.16 arcsec (bottom left bars).}
\label{FIG:HD169142_profiles}
\end{figure}

\subsection{Gas surface density}
\label{SUBSEC:Gas_surface_density}
The gas surface density can be inferred using both an optically thin and an optically thick line. We use the $J: 2\rightarrow 1$  lines and assume that  $^{12}$CO is optically thick and the  $^{13}$CO is optically thin. The $^{13}$CO column density for the $J: 2\rightarrow 1$  rotational  transition (e.g. \cite{Estalella_1994}) is given by
\begin{equation}
N^{13\rm CO} = c_0 \tau_0^{13} T_{\rm ex} \Delta v \left[ \frac{\exp[3h\nu/(2k_BT_{\rm ex})] }{\exp \left[ h\nu/(k_BT_{\rm ex}) \right] - 1}   \right],
\label{Eq:13CO_column density}
\end{equation}
 where $c_0= 16 \pi k_B\nu^2 J /[(2J+1)hc^3A_{J,J-1}]$, $h,k_B$ are the Planck and Boltzmann constants, $\Delta v$ is the line width and $\nu=230$ GHz, $A_{2,1} = 3.25 \times 10^{-7} \ \rm{s}^{-1}$ are the frequency and the Einstein coefficient for this transition, respectively. The excitation temperature  $T_{\rm ex}$ is given by
\begin{equation}
T_{\rm ex} = \frac{h\nu/k_B}{\ln \left[1 + \frac{h\nu/k_B}{T_0^{12} + {\cal J}^{12}(2.7 \rm K)} \right]},
\label{Eq:Excitation_temperature}
\end{equation}
where  $T_0^{12}$ is the brightness temperature of the $^{12}$CO  J: $2 \rightarrow 1$ line, ${\cal J}_\nu (T)= (h \nu/k) /  (\exp(h\nu/kT -1)$ is the intensity in units of temperature, and
 ${\cal J}^{12}(2.7 \rm K) $ is the intensity  at the frequency of the $^{12}$ CO { J: 2 $\rightarrow$1} evaluated at the background temperature. 
 Finally, the optical depth of the $^{13}$CO J: 2 $\rightarrow$ 1 line ($\tau_0 ^{13}$)  is given by
\begin{equation}
\tau_0^{13} = - \ln \left[1 - \frac{T_0^{13}}{{\cal J}^{13}(T_{\rm ex}) - {\cal J}^{13}(2.7 \rm K)} \right],
\end{equation}
where $T_0^{13}$ is the brightness temperature of the $^{13}$CO J: 2 $\rightarrow$ 1 line, and ${\cal J}^{13}(T_{\rm ex})$ is the intensity evaluated at the excitation temperature. We find that $\tau_0^{13} < 0.8$ throughout the disk.
 
To obtain the gas surface density at each radius one needs an abundance factor between $^{13}$CO and $\rm H_2$ (which dominates the gas mass), such that
$\Sigma_{\rm g} = m_{\rm H_2} [\rm H_2 /^{13}CO]  \times N^{13\rm CO}$. The abundance is obtained by normalizing the gas surface density with the total mass of the disk 
$M_{\rm disk} = 2\pi \int_0^{R} \Sigma_{\rm g} \varpi d\varpi$.
For a disk mass of $M_{\rm disk} =  0.019 M_{\odot}$ \citep{Fedele_2017},
the abundance between the $^{13}$CO and the $\rm H_2$ molecules of [$ ^{13}$CO/$\rm H_2$] = $1.0 \times 10^{-5}$, similar to values found in the ISM (e.g. \cite{Dickman_1978}). 

The top panel of Figure (\ref{FIG:Disk_properties}) shows the gas surface density model as a function of the radius. The width of the line gives the uncertainty in the gas surface density due to the noise of the CO maps and the error propagation. 
In the next section we use this gas surface density to model the dust surface density using Equation (\ref{EQ:Total_dust_surface_density}). 
\begin{figure}[!t]
\centering
\includegraphics[scale=0.6]{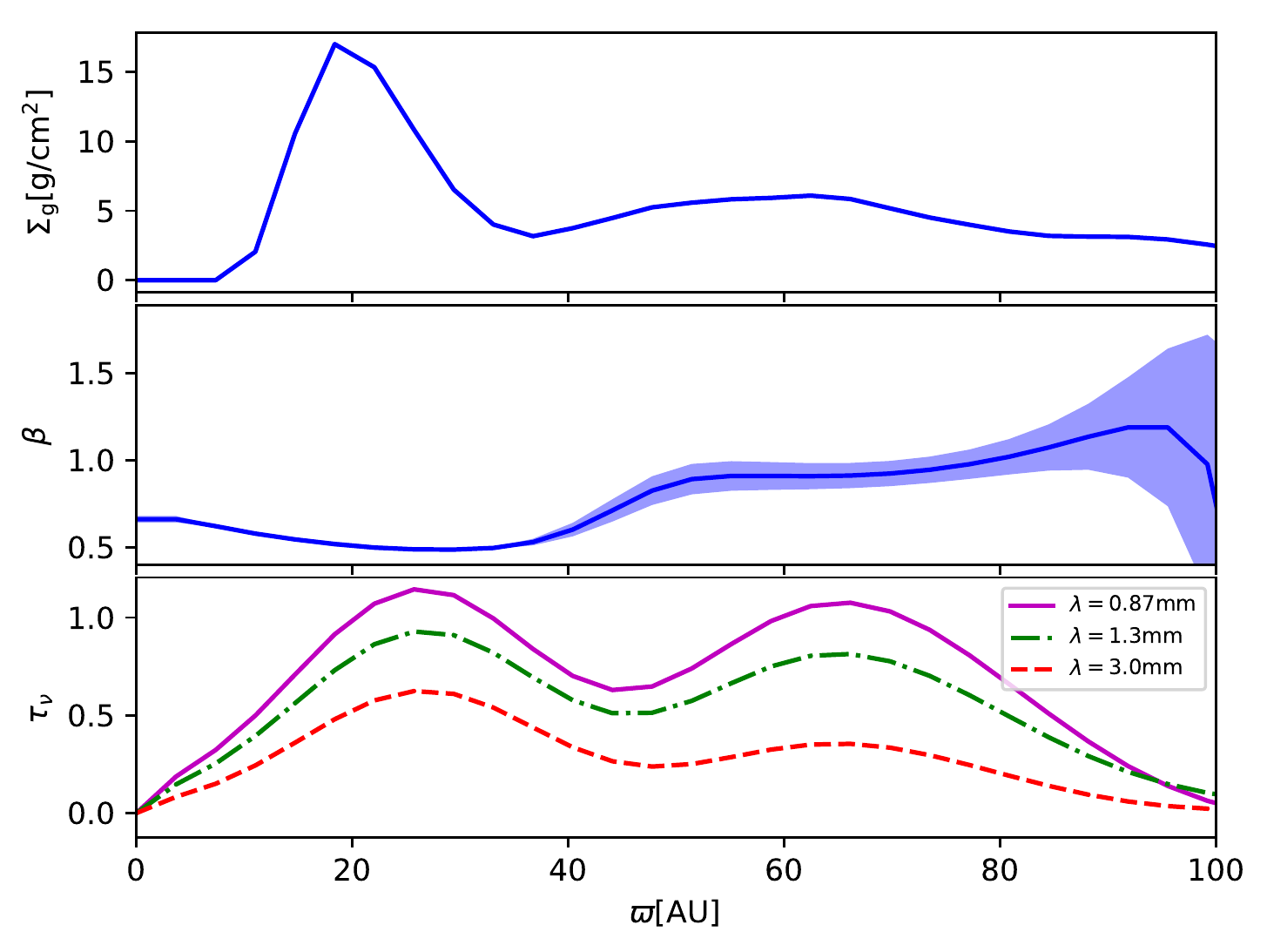}
\caption{HD 169142 disk properties. Top panel: Gas surface density model derived from the assumed $^{12}$CO (optically thick) and $^{13}$CO (optically thin) emission. Middle panel: Spectral index $\beta$ (between $\lambda = 870 \mu$m and $3$ mm) of the dust emission . Bottom panel: Optical depth at $\lambda = 0.87$ mm (magenta solid line), 1.3 mm (green dash-dotted line), and 3.0 mm (red dashed line).}
\label{FIG:Disk_properties}
\end{figure}
 
\vspace{1cm}
\subsection{Dust properties and radiative transfer model}
\label{SUBSEC:Radiative_transfer}
The emergent specific intensity is given by 
\begin{equation}
I_{\nu} = I_{\nu}^{\rm bg} e^{-\tau_{\nu}} + \int_0^{\tau_{\nu}} S_{\nu}(T) e^{-t}dt
\label{EQ:Intensity}
\end{equation}
where $I_{\nu}^{\rm bg}$ is the background intensity, $\tau_{\nu} = (\kappa_\nu + \sigma_\nu) \Sigma_d$ is the optical depth at the frequency $\nu$, 
where $\kappa_\nu$ is the mass absorption coefficient and $\sigma_\nu$ is the mass scattering coefficient.
The source function $S_{\nu}(T)$ is  given by \citep{Mihalas_1978}
\begin{equation}
S_{\nu}(T) = \omega_{\nu}J_{\nu} + (1-\omega_{\nu})B_{\nu}(T),
\label{EQ:Source_function}
\end{equation}
where the albedo is $\omega_{\nu} = \sigma_\nu/ (\kappa_\nu + \sigma_\nu)$,  $B_{\nu}(T)$ is the Planck function, and $J_{\nu}$ is the local mean intensity. We approximate $J_{\nu}$ by the analytical solution found by \cite{Miyake_1993} (hereafter MI93) for a vertically isothermal slab. Then, the source function can be written as
\begin{equation}
S_{\nu}(T) = B_{\nu}(T) \left[ 1 + \omega_{\nu} f(t, \tau_{\nu}, \omega_{\nu}) \right],
\label{EQ:Source_sim1}
\end{equation}
where
\begin{equation}
f(t, \tau_{\nu}, \omega_{\nu}) = \frac{\exp(-\sqrt{3}\epsilon_{\nu}t)   + \exp(\sqrt{3}\epsilon_{\nu}(t - \tau_{\nu}))}{\exp(-\sqrt{3}\epsilon_{\nu}\tau_{\nu}) (\epsilon_{\nu}-1) - (\epsilon_{\nu}+1)},
\label{EQ:Source_sim2}
\end{equation}
$\tau_\nu$ is the total optical depth and $t$ is the variable optical depth, both measured perpendicular to the disk mid plane, and
\begin{equation}
\epsilon_{\nu} = \sqrt{1-\omega_{\nu}}.
\label{EQ:Source_sim3}
\end{equation}
For a face-on and vertically isothermal disk, the solution of  Equation (\ref{EQ:Intensity}) using Equations (\ref{EQ:Source_sim1})-(\ref{EQ:Source_sim3}) and setting $I_{\nu}^{\rm bg} = 0$ is
\begin{equation} 
I_{\nu} = B_{\nu}(T) \left[ \left(1 - \exp(-\tau_{\nu}) \right) + \omega_{\nu}\cal F(\tau_{\nu}, \omega_{\nu}) \right],
\label{EQ:RAD_General}
\end{equation}
where
\begin{eqnarray}
\nonumber \cal F(\tau_{\nu}, \omega_{\nu}) &=& \frac{1}{\exp(-\sqrt{3}\epsilon_{\nu}\tau_{\nu}) (\epsilon_{\nu}-1) - (\epsilon_{\nu}+1)} \times \\
 \nonumber && \left[ \frac{1 - \exp(-(\sqrt{3}\epsilon_{\nu}+1)\tau_{\nu})}{\sqrt{3}\epsilon_{\nu} + 1} \right. \\
 && \left. + \frac{\exp(-\tau_{\nu}) - \exp(-\sqrt{3}\epsilon_{\nu} \tau_{\nu}) }{\sqrt{3}\epsilon_{\nu} - 1}\right].
 \label{EQ:F_int}
\end{eqnarray}
Then, in the optically thin regime, the emergent intensity is
\begin{equation}
I_{\nu}^{\rm thin} = B_{\nu}(T) \tau_{\nu}(1 - \omega_{\nu}) = B_{\nu}(T) \kappa_\nu \Sigma_{\rm d} ,
\label{EQ:Thin}
\end{equation}
independent of the albedo (see MI93).  
In the optically thick regime,  the emergent intensity is
\begin{equation}
I_{\nu}^{\rm thick}=B_{\nu}(T) \left[ 1 - \frac{\omega_{\nu}}{ (\epsilon_{\nu}+1)(\sqrt{3}\epsilon_{\nu} +1 )    }   \right],
\label{EQ:Thick}
\end{equation}
consistent with the discussion in \cite{Dalessio_2001} (hereafter DA01).

We assume a power-law dust temperature 
\begin{equation}
T_{\rm d} = T_0 \left(\frac{\varpi}{\varpi_0} \right)^{-q},
\end{equation}
where $T_0 = 500$ K at $\varpi_0 = 1$ AU and $q = 1/2$. This slope is estimated by assuming an equilibrium between the star incident radiation and the dust grains emission, such that $T_{\rm d}^4 \approx  T_*^4  W(\varpi)$, where $W(\varpi)$ is the dilution factor given by $W(\varpi) =  1/2 (1 - [1-(R_*/\varpi)^2]^{1/2})$, and $T_*, R_*$ are the  effective temperature and radius of the star respectively (Subsection \ref{SUBSEC:Observations}). Since a large dust central hole has been previously reported in the disk around HD 169142 (e.g. \citealt{Fedele_2017}, \citealt{Macias_2017}), the inner and outer dust radius are set to be 15 AU and 90 AU respectively.

The monochromatic opacity ($\chi_{\nu} = \kappa_{\nu} + \sigma_{\nu}$) and albedo are computed with the code of (DA01) for a dust composition of silicates, organics, and ice with the relative abundances described by \cite{Pollack_1994}. In this code, the total dust opacity is given by
\begin{equation}
\chi_{\nu} = \frac{\int_{\rm a_{\rm min}}^{a_{\rm max}(\varpi)} \chi_{\nu}(a)n(a)da}{\int_{\rm a_{\rm min}}^{a_{\rm max}(\varpi)} n(a)da },
\end{equation}
where $\chi_{\nu}(a)$ is the opacity of each dust grain size at frequency $\nu$, and the particle size distribution has $n(a)da \propto a^{-p}da$ with $p = 3.5$,  the minimum grain size is $a_{\rm min} = 0.05\mu$m, and the maximum grain size $a_{\rm max}(\varpi)$ depends on the disk radius. The opacity and albedo curves as a function of the wavelength for different dust size distributions are shown in Figures 9 and 10 of  \cite{Sierra_2017}. 

To determine  the maximum grain size, $a_{\rm max}(\varpi)$, one has to fit a power-law to the observed intensity $I_\nu$. This 
comes from the following considerations:  in the optically thin limit and, including only the mass absorption coefficient  $\kappa_\nu$, $I_{\nu} \propto \kappa_{\nu} B_{\nu}(T)$. Also, at low frequencies, the Planck function can be approximated as $B_{\nu}(T) =\nu^2 k T/2c^2 $ (Rayleigh-Jeans regime), and the opacity is given by a  power-law of the frequency $\kappa_{\nu} \propto \nu^{\beta}$ \citep{Beckwith_1990}. Then the specific intensity will follow a power-law of the  frequency as $I_{\nu} \propto \nu^{2+\beta}$. In 
general, $\beta$ will be a function of radius. Since, for a given slope $p$ of the dust size distribution, $\beta$ depends on the maximum grain radius (e.g., \cite{Ossenkopf_1994, Pollack_1994}), one can find $a_{\rm max} (\varpi)$ at each radius \footnote{See discussion about on the degeneracy between $a_{\rm max}$ and the slope $p$ in Section \ref{SEC:amax_p_degeneration}.}.
Nevertheless, one can avoid the assumptions of optically thin emission and  the Rayleigh-Jeans regime, and include the scattering mass opacity $\sigma_{\nu}$ in the total opacity, $\chi_\nu = \kappa_\nu + \sigma_\nu$. In this case, 
the parameter $\beta$ can be derived from the optical depth that appears in Equation (\ref{EQ:RAD_General}), $\tau_{\nu} = \tau_0 (\nu/\nu_0)^{\beta}$, where $\tau_0$ is the optical depth at a reference frequency $\nu_0$. 

The middle panel of Figure (\ref{FIG:Disk_properties}) shows $\beta$ derived from the dust continuum observations following the latter procedure, without assuming optically thin emission nor the Rayleigh-Jeans approximation, and including the mass scattering opacity. The width of the curve represents the error in the fit. The error is negligible for $\varpi < 80$ AU, where the signal-to-noise ratio in the dust emission maps is high. The bottom panel of the same Figure shows the optical depth at $\lambda = 870\ \mu$m (magenta solid line), 1.3 mm (green das-dotted line), and 3.0 mm (red dashed lin): the disk is optically thin at $\lambda = 1.3$mm and $3.0$ mm; it is only marginally thick close to the two maxima at $\lambda = 870\ \mu$m.

Figure (\ref{FIG:amax}) shows the maximum grain size, $a_{\mathrm{max}}(\varpi)$, as a function of the disk radius assuming a slope of the particle size distribution of $p = 2.5$ (magenta solid line), $3.0$ (green dash-dotted line), and $3.5$ (red dashed line) (see more details in Appendix \ref{APP:Beta}). In the following we assume $p=3.5$ and in Section (\ref{SEC:amax_p_degeneration}) we discuss this assumption.

We find that $a_{\rm max}(\varpi)$ is very large at the position of the first maximum and decreases with the disk radius, reaching a value of $\sim 2$ mm at $\varpi = 90$ AU. This behavior could be due to differential radial migration of the dust grains toward the gas pressure maxima and/or dust growth.

\begin{figure}
\centering
\includegraphics[scale=0.5]{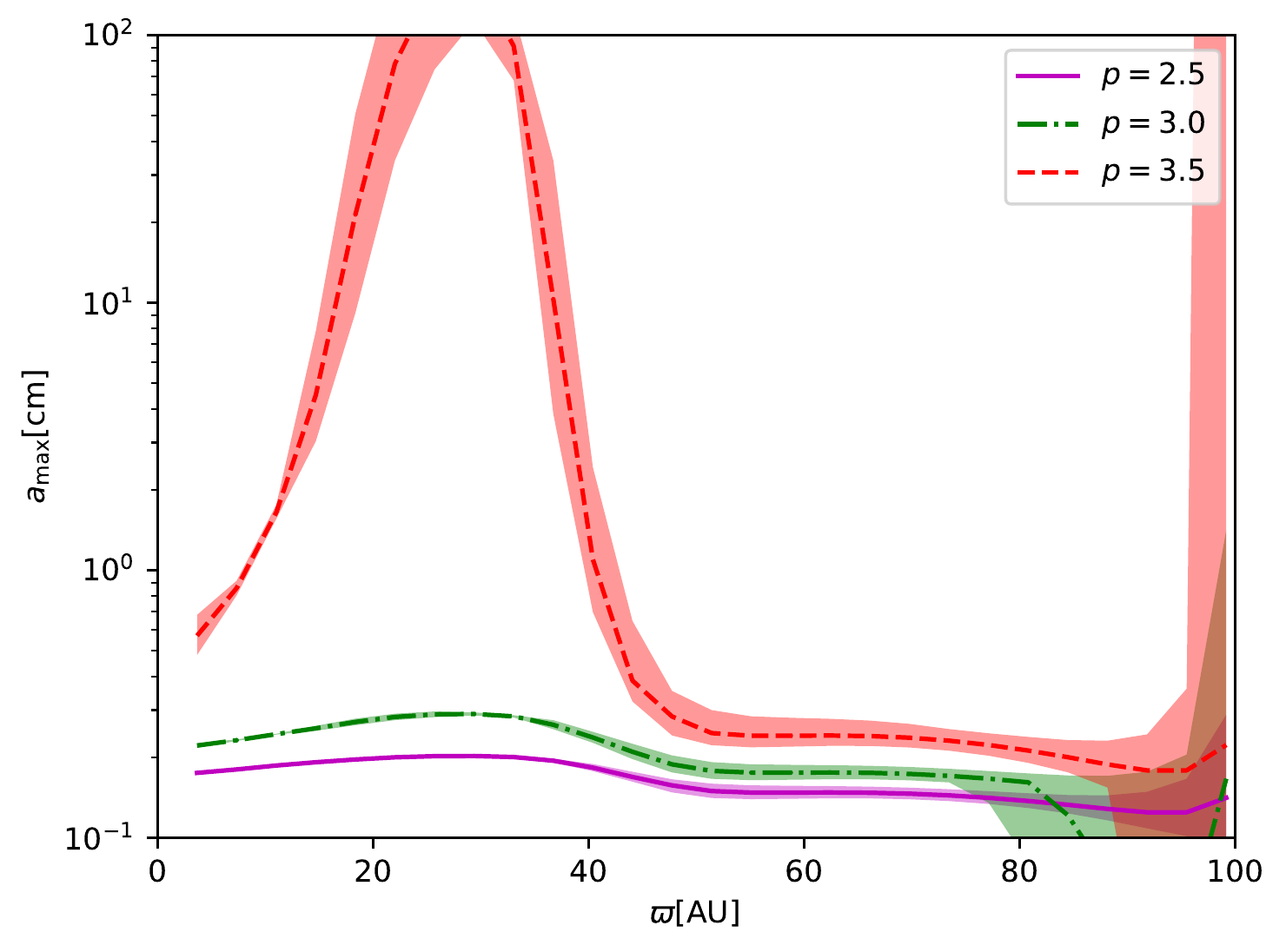}
\caption{Maximum grain size as a function of the radius for different slopes in the particle size distribution: $p = 2.5$ (magenta solid line), $3.0$ (green dash-dotted line), and $3.5$ (red dashed line).}
 \label{FIG:amax}
\end{figure}

\subsection{Best fit dust model}
\label{SUBSEC:Model_grid}
The dust-to-gas mass ratio $\epsilon_{\rm global}$ and the $\alpha$ parameter that appear in Equations (\ref{EQ:k}) and (\ref{EQ:Total_dust_surface_density}) are the unknown parameters of the dust model.

These parameters are fitted by creating a grid of models and comparing their radial intensity profiles ($I^{\rm mod}_\lambda (\varpi)$) with the observed azimuthally averaged intensity profiles ($I^{\rm obs}_\lambda (\varpi)$) at each wavelength. We vary $\alpha$  from $10^{-5}$ to $10^{-1}$, while $\epsilon_{\rm global}$ varies from $10^{-4}$ to $10^{-1}$. The number of models for each parameter is $50$ (the total number of models is $50\times 50$) and the values are logarithmically equally spaced.
The top panels of Figure (\ref{FIG:Parameter_space}) show the reduced chi-squared 
\begin{equation}
\chi^2_{r,\lambda} = \frac{1}{N_p} \sum_{\varpi} \left[\frac{\rm I^{\rm mod}_\lambda (\varpi) - I^{\rm obs}_\lambda (\varpi)}{\sigma ^{\rm obs}_\lambda(\varpi)} \right] ^2,
\end{equation}
of the dust models at $\lambda = 870\ \mu$m (left panel), $1.3$ mm (middle panel), and $3.0$ mm (right panel), where $N_p$ is the number of radii and $\sigma ^{\rm obs}_\lambda(\varpi)$ is the uncertainty of the observed intensity at each radius. The latter is given by 
\begin{equation}
\sigma ^{\rm obs}_\lambda(\varpi) = \rm{RMS_\lambda}/\sqrt{{\it n}},
\end{equation}
where RMS$_{\lambda}$ is the root mean square noise of the observed map, and $n$ is the number of beams within the area of each ring. 
The bottom panels show the intensity profiles of the best dust models (red solid line), the optical depth models (green dashed line), and the observed intensity profiles (blue line) whose error is given by the width of the line.

\begin{figure*}[!t]
\centering
\includegraphics[scale=0.4]{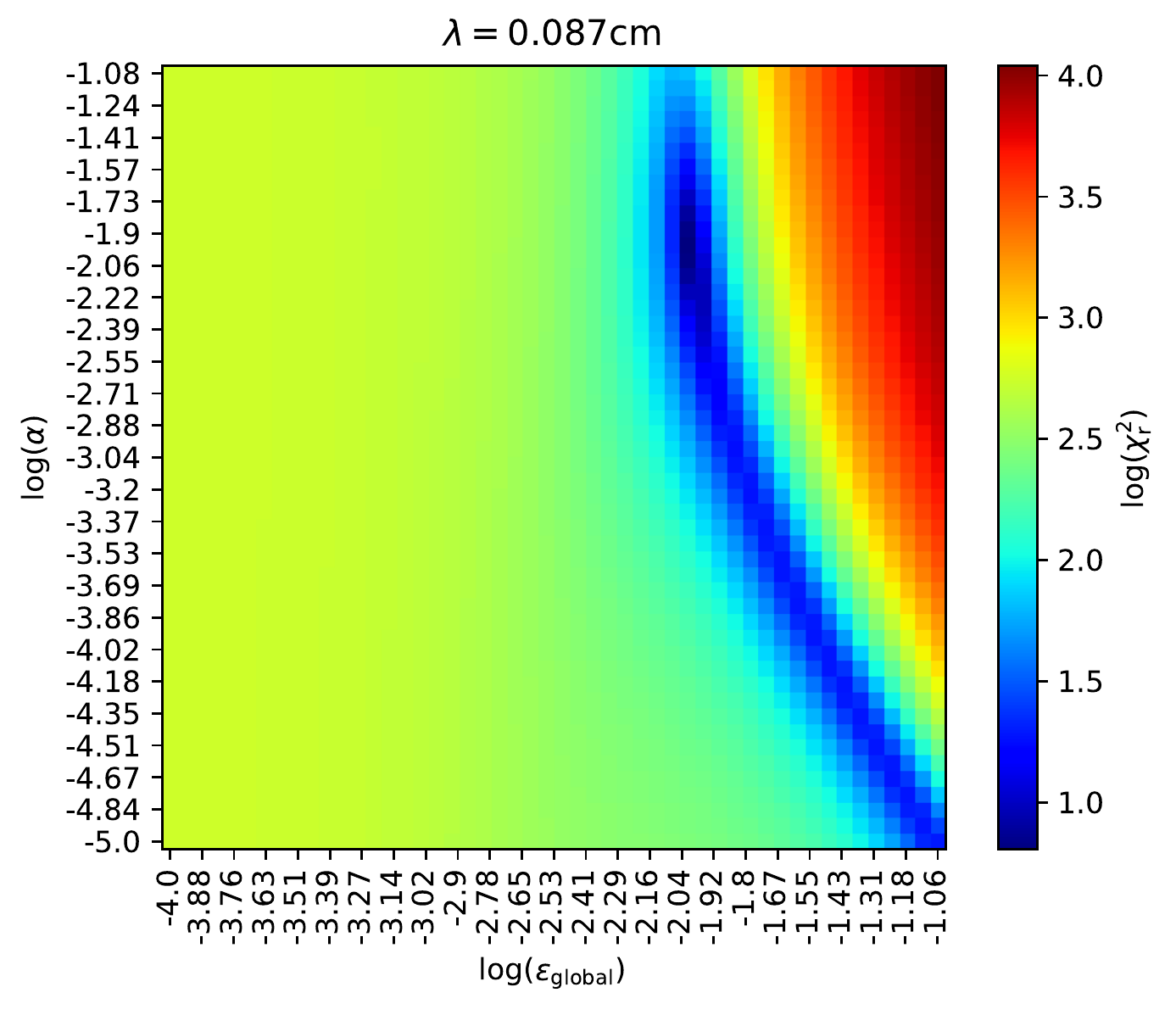} 
\includegraphics[scale=0.4]{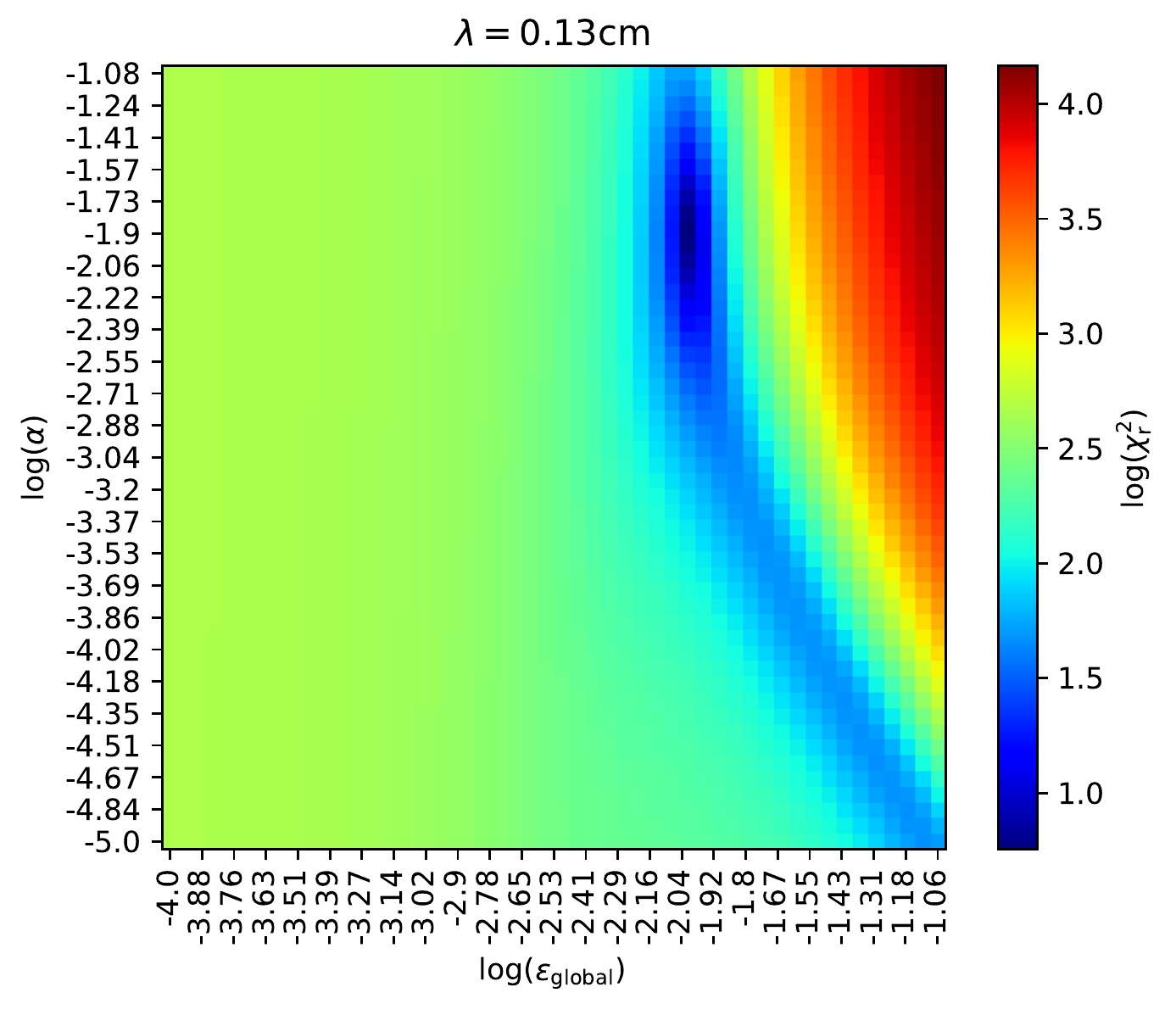} 
\includegraphics[scale=0.4]{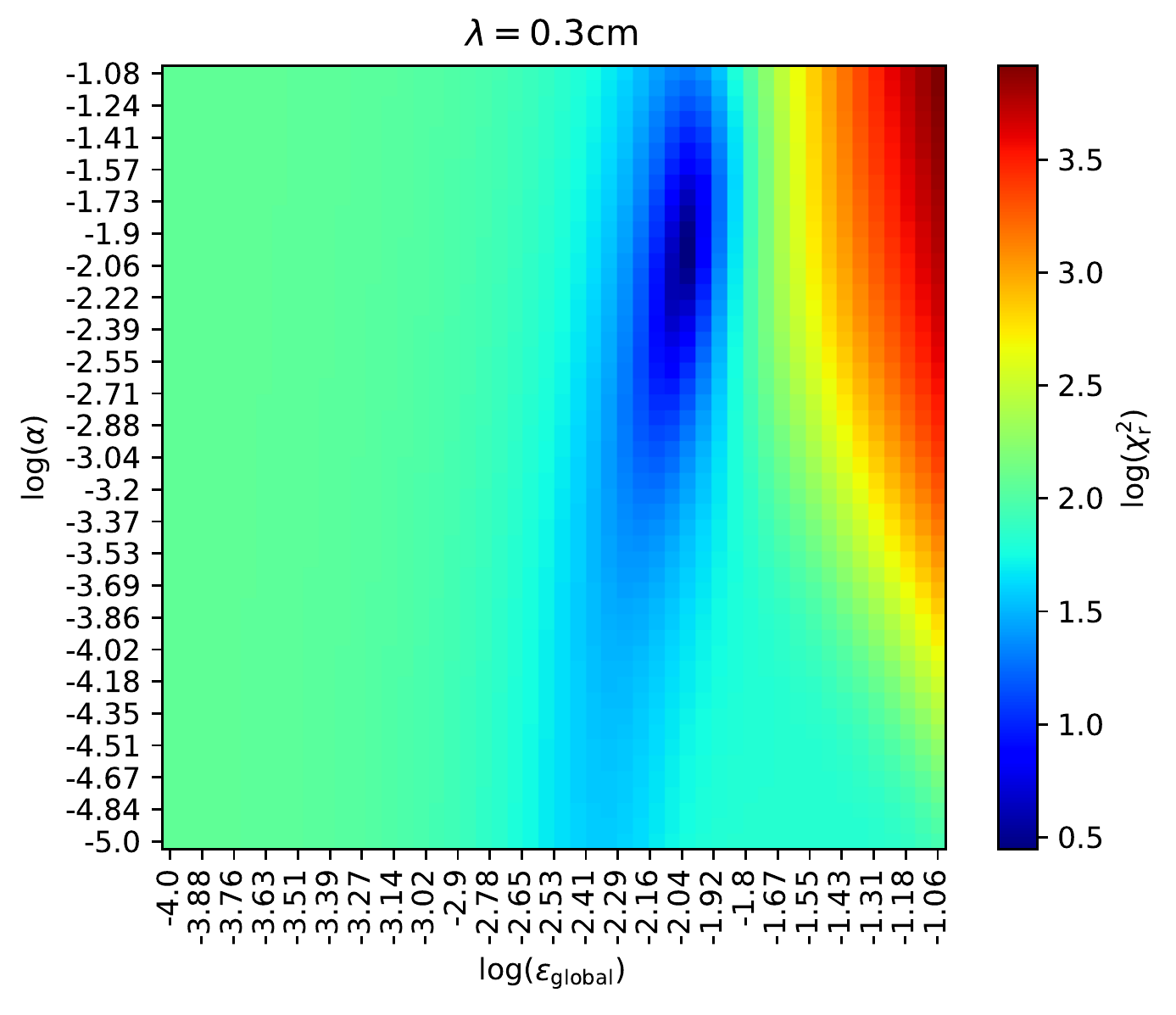} 
\includegraphics[scale=0.35]{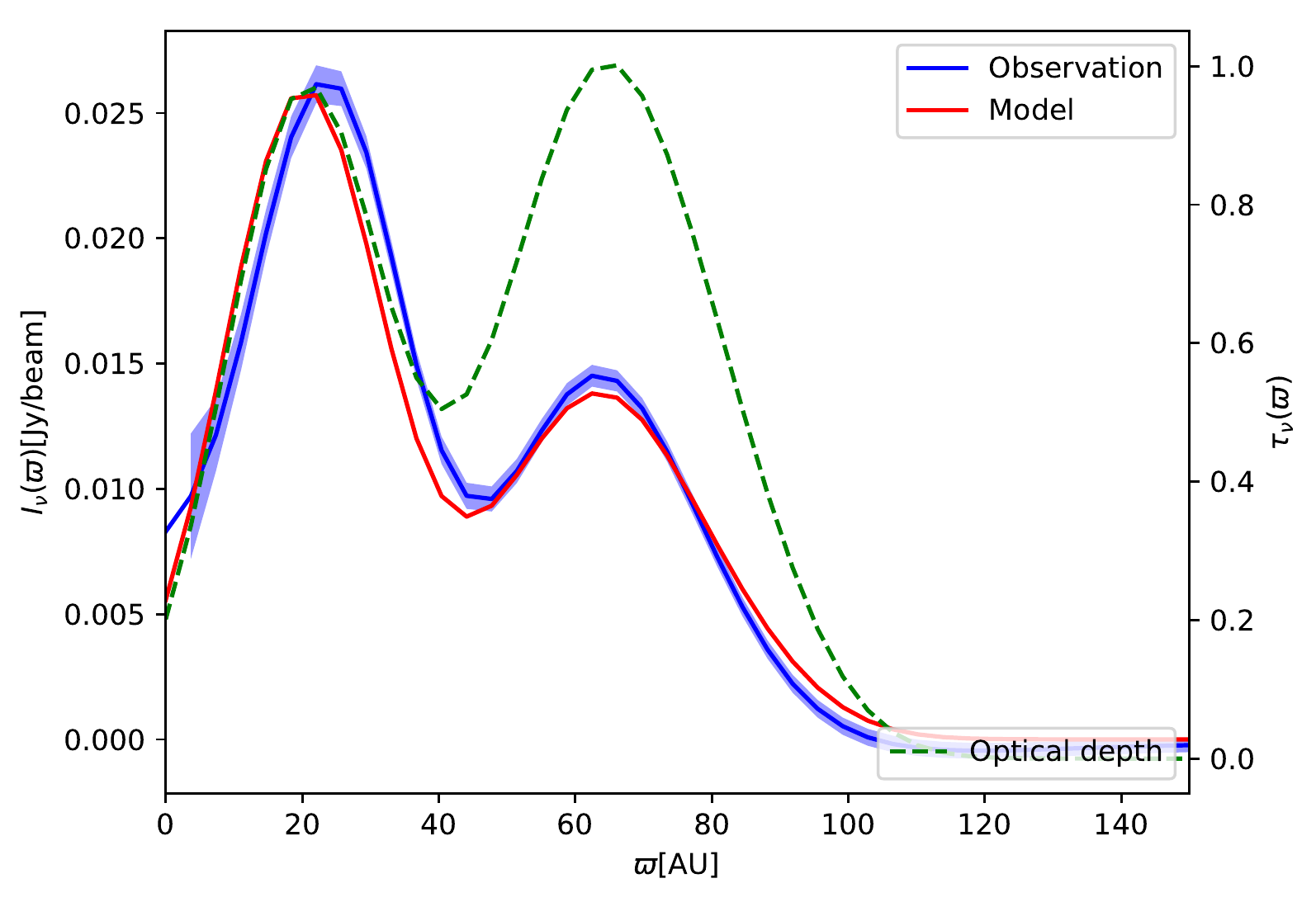} 
\includegraphics[scale=0.35]{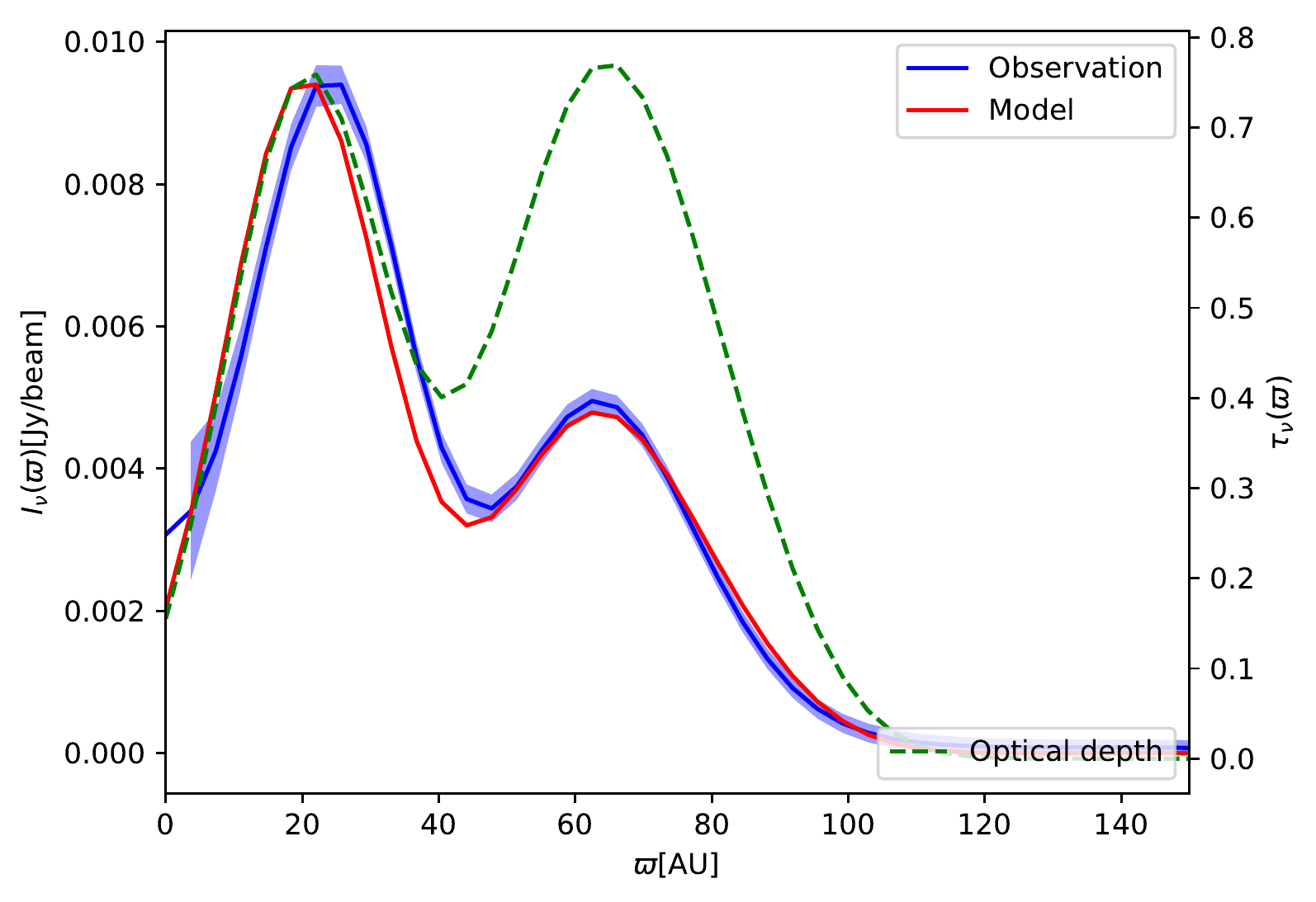} 
\includegraphics[scale=0.35]{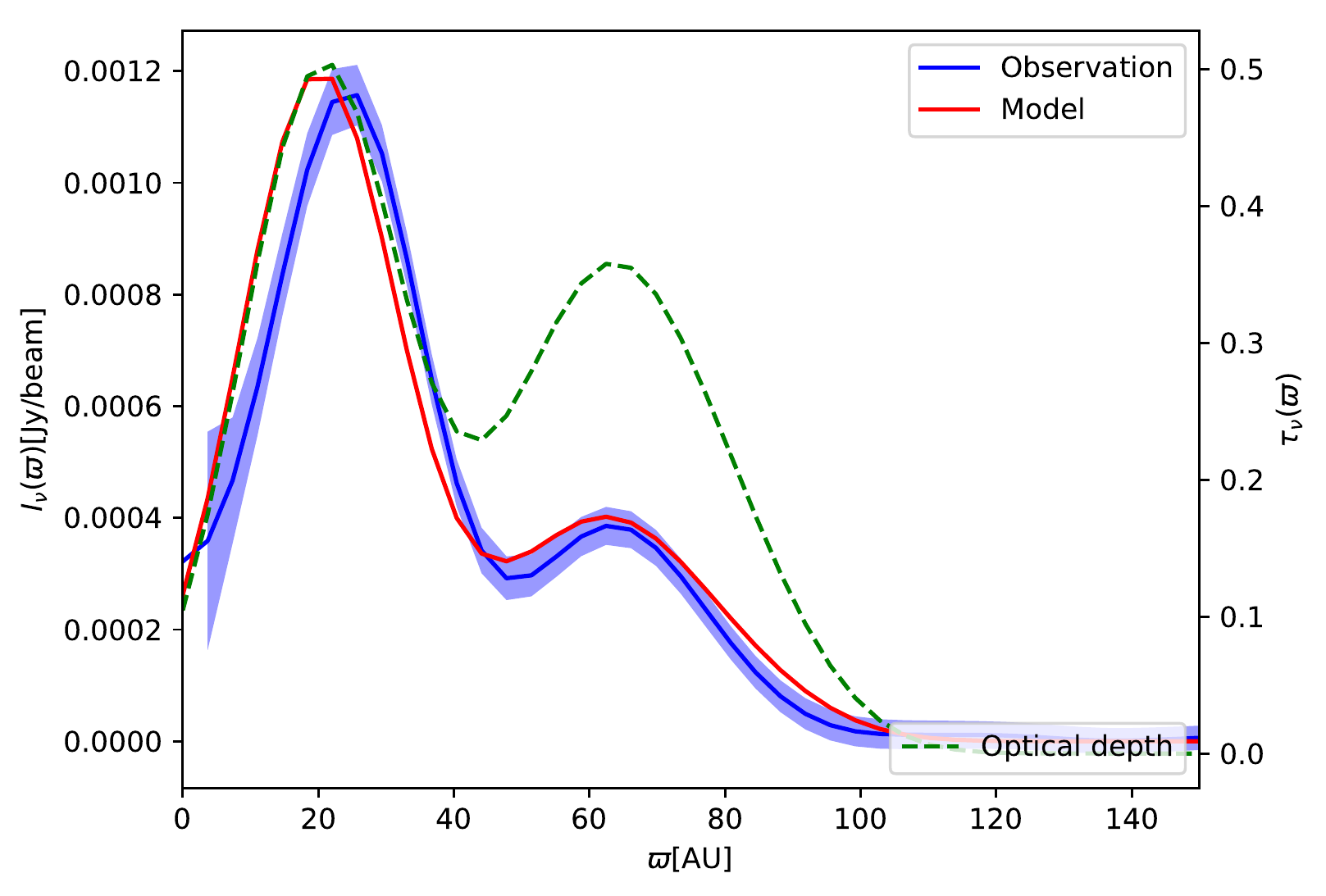} 
\caption{Exploration of the parameter space for the dust continuum emission at $\lambda = 870\ \mu$m (left panels), $1.3$ mm (middle panels), and $3.0$ mm (right panels). Top panels: Reduced chi-squared ($\chi_r^2$) for different dust models in the ($\epsilon_{\rm global}, \alpha$) space. Bottom panels: Best model at each wavelength; the blue line is the observational dust continuum emission, the red solid line is the best emission model and the green dashed line is the best optical depth model.}
\label{FIG:Parameter_space}
\end{figure*}

Figure (\ref{FIG:Best_fits}) plots the isocontours where the value of the reduced chi-squared is 1.5 times the minimum value of $\chi^2_{r,\lambda}$ for all wavelengths: $\lambda = 870\ \mu$m (red solid line), $1.3$ mm (green dashed line), and $3.0$ mm (blue dash-dotted line). The best parameters at each wavelength are summarized in Table (\ref{TAB:Best_parameters}). To obtain a global model, one requires that the parameters do not depend on wavelength. The best values for $\epsilon_{\rm global}$ are the same, and 
the best values of $\alpha$ slightly differ with wavelength. Therefore, the best averaged parameters that describe the dust concentration in the disk of HD 169142 are: $\epsilon_{\rm global} = 1.05 \times 10^{-2}$ 
and $\alpha = 1.35\times 10^{-2}$; this means that the total dust mass is $2.0 \times 10^{-4} \ M_{\odot}$.
\begin{figure}[!t]
\centering
\includegraphics[scale=0.5]{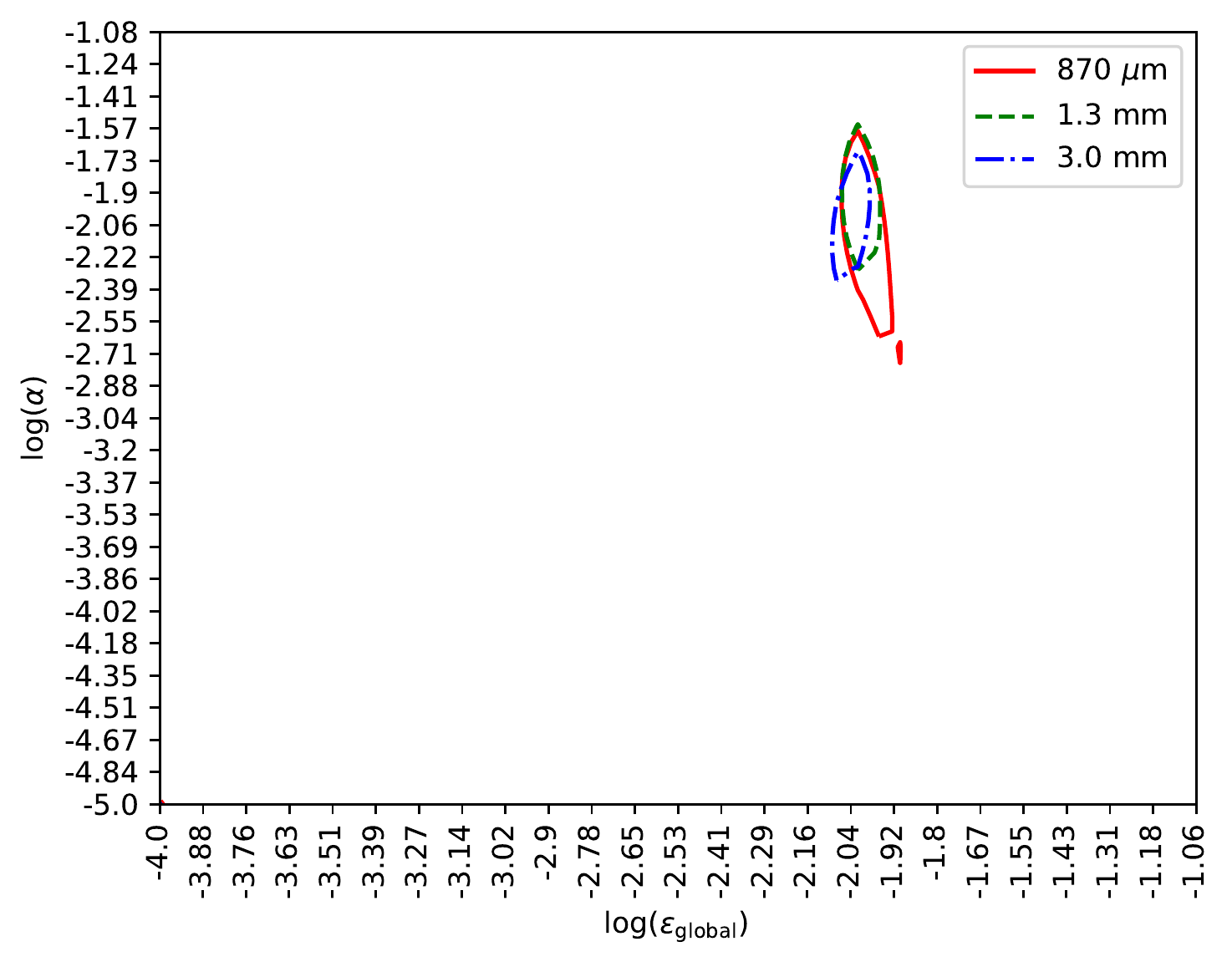}
\caption{Isocontours where the reduced chi-squared is equal to 1.5 times the minimum value of $\chi_r^2$ for each wavelength $\lambda = 870\ \mu$m (red solid line), $1.3$ mm (green dashed line), and $3.0$ mm (blue dash-dotted line).}
\label{FIG:Best_fits}
\end{figure}

Note that the fitted value of $\alpha$ depends on the resolution of the gas structure. For example, if in higher angular resolution observations a ring fragments into thinner rings, that would require a smaller value of $\alpha$ to concentrate the dust in these rings.

\begin{table}[t!]
\centering
\caption{Best physical parameters}
\begin{tabular}{ccc}
\hline 
$\lambda$ (mm) & $\alpha/10^{-2}$ & $\epsilon_{\rm global}/10^{-2}$ \\
\hline
0.87   & 1.26 & 1.05 \\
1.3     & 1.53 & 1.05 \\
3.0     & 1.26 & 1.05 \\
\hline
\end{tabular}
\label{TAB:Best_parameters}
\end{table}

Figure (\ref{FIG:Best_maps}) shows the disk maps for different wavelengths; the left panels are the observed maps, the middle panels show the emission of the dust model with the best averaged parameters, 
and the right panels are the maps of 
absolute differences ($\vert  {\rm observation - model} \vert$). The wavelength increases from top to bottom ($\lambda = 870\ \mu$m, $1.3$ mm, $3$ mm respectively). Note that the main difference between the observations and model are the non-axisymmetric structures, which are not taken into account in the dust model.

\begin{figure*}[!t]
\centering
\includegraphics[scale=1]{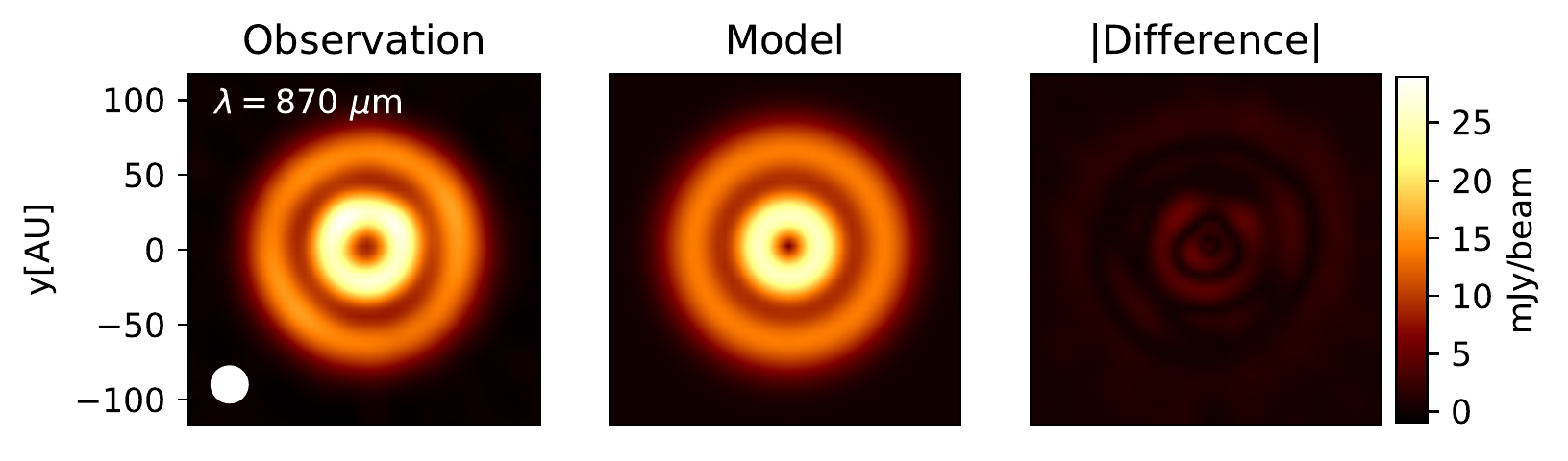} \includegraphics[scale=1]{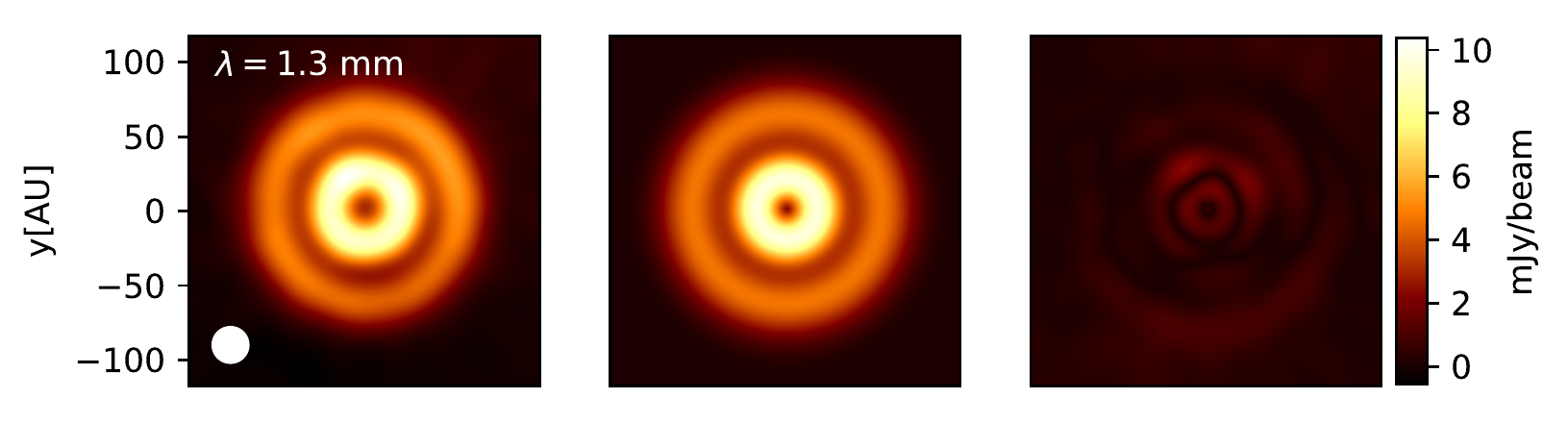} 
\includegraphics[scale=1]{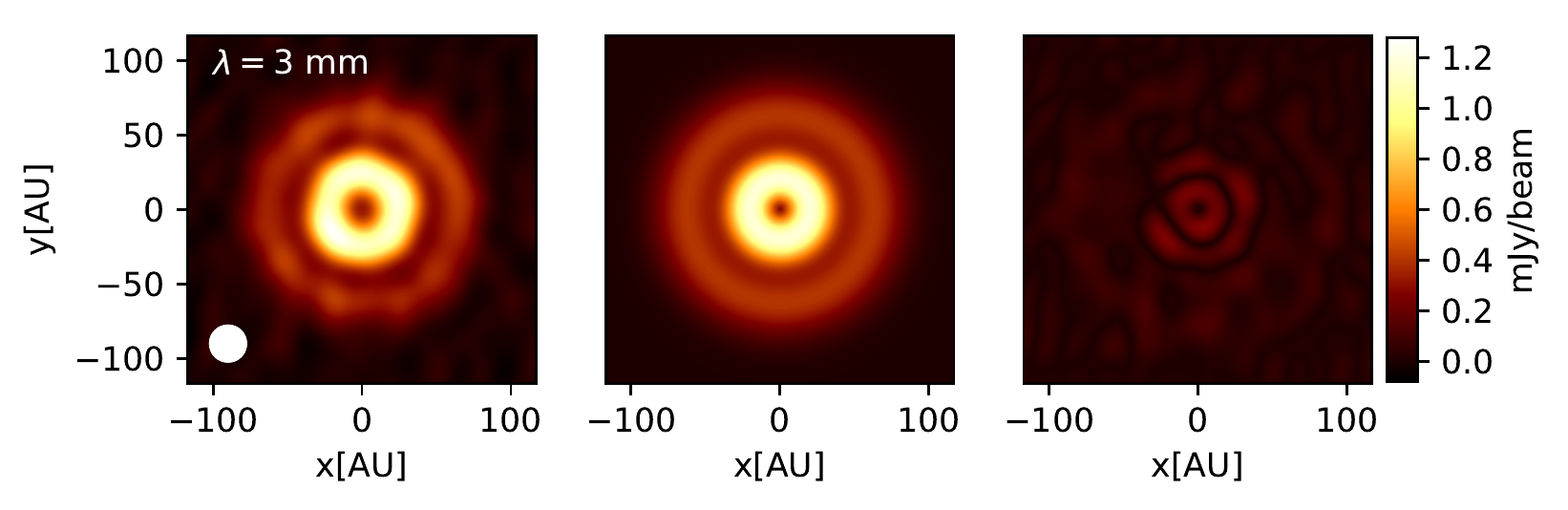}
\caption{Maps of the thermal dust emission at $\lambda = 870\ \mu$m (first row), $1.3$ mm (second row), and $3.0$ mm (third row). The left, middle and right panels are the observational, model and difference maps.}
\label{FIG:Best_maps}
\end{figure*}
 
For the best averaged parameters, Figure (\ref{FIG:Dust_properties}) shows the dust surface density (the red solid line), the local dust-to-gas mass ratio (green dashed line), the global dust-to-gas mass ratio $\epsilon_{\rm global}$ (green dotted line), and the gas surface density scaled by the global dust-to-gas mass ratio (blue dash-dotted line). The latter would mimic the dust surface density if there is no dust migration, which is not the case in HD 169142. The local dust-to-gas mass ratio in the first maximum increases by a factor of $6$ compared with the global dust-to-gas mass ratio; while in the second local maximum it is below the average value. In addition, this figure shows that the maximum dust-to-gas mass ratio is less than 0.06, therefore, one can neglect the back reaction of the dust on the gas.

The amplitude of the dust surface density in the second maximum ($\varpi = 60$ AU) is much smaller than in the first maximum ($\varpi = 24$ AU), in contrast with the behaviour of the intensity profiles (bottom panels of Figure (\ref{FIG:Parameter_space})); this occurs because of opacity effects associated to the maximum grain size at each radius: In the first maximum, the grains have a maximum size  larger than $10$ cm;  such grains do not have a large opacity at millimeter wavelengths. However, in the second maximum, the grains have a maximum size $\sim 2$ mm, which have a large opacity at millimeter wavelengths. For this reason, the contrast in the maxima of the dust surface density is larger than expected. Appendix \ref{APP:Beta} explains the effect of the maximum grain size in the opacity for different slopes of the particle size distribution. 

\begin{figure}[!t]
\centering
\includegraphics[scale=0.5]{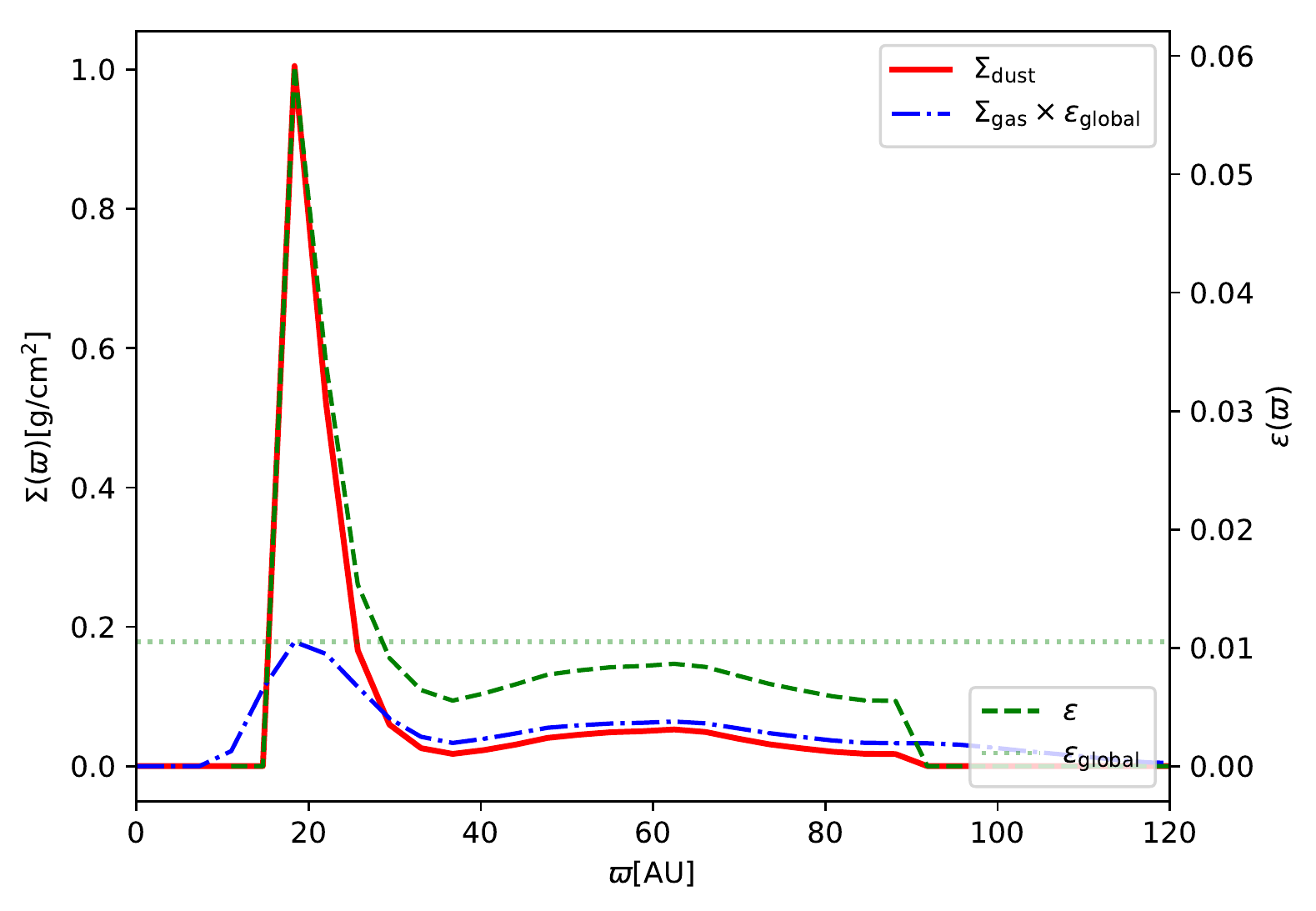}
\caption{Dust properties using the best averaged parameters: $\epsilon_{\rm global} = 1.05\times 10^{-2} $, $\alpha = 1.35\times 10^{-2}$. The red solid line is the dust surface density, the blue dash-dotted line is the gas surface density scaled by the global dust-to-gas mass ratio, the green dashed line is the local dust-to-gas mass ratio, and the green dotted line is the global dust-to-gas mass ratio.}
\label{FIG:Dust_properties}
\end{figure}

\section{$a_{\rm max}, p$ degeneration}
\label{SEC:amax_p_degeneration}
In this section we study the local changes of the particle size distribution due to dust size differential  migration. This effect has been already found by e.g. \cite{Pinte_2016}, \cite{Sierra_2017}.

The model to obtain the slope $p$ as a function of the disk radius based on the results of the analytical model (Section \ref{SEC:Analytical_model}) as follows: If the total dust surface density is given by  Equation (\ref{EQ:Total_dust_density}), the term $\Sigma_{\rm d} (\varpi,a)n(a) d a$ can be interpreted as proportional to the local particle size distribution after dust concentration. If $n_2(a) da $ is this new distribution and $c_1$ the proportionality factor, then,
\begin{equation}
 n_2(a) da = c_1 \Sigma_{\rm d}(\varpi,a) n(a) da.
\label{EQ:n2_nonorm}
\end{equation}
The number of particles per unit volume must be the same as the original when averaged in all the disk area $A$; then $\int_A n_2(a) da \, dA = \int_A n(a) da \, dA $. So, the proportionality factor is constrained by
\begin{equation}
c_1 =  \frac{A}{\int_A \Sigma_{\rm d}(\varpi,a) dA}.
\label{EQ:n2_const}
\end{equation}
Then, substituting  Equation (\ref{EQ:Dust_model}) and  Equation (\ref{EQ:n2_const}) in Equation (\ref{EQ:n2_nonorm}), one obtains
\begin{eqnarray}
\nonumber n_2(a) d a &=& \left[\frac{A \Sigma_{\rm g}(\varpi) \exp[-ka]}{\int_A \Sigma_{\rm g}(\varpi) \exp[-ka]dA}\right] n(a) da ,\\
a^{-p_2} &\propto & \left[ \frac{\exp[-ka]}{\int_A \Sigma_{\rm g}(\varpi) \exp[-ka]dA}\right] a^{-p}.
\label{EQ:n2}
\end{eqnarray}
Equation (\ref{EQ:n2}) only depends on the viscosity coefficient $\alpha$ via the factor $k$ defined in Equation (\ref{EQ:k}). One expects that, if the viscosity coefficient $\alpha \rightarrow \infty$ (i.e. $k \rightarrow 0$), there is no dust differential migration and the dust particle size distribution does not change, i.e., $p_2 = p$. 

For the gas surface density $\Sigma_{\rm g}(\varpi)$ in the top panel of Figure (\ref{FIG:Disk_properties}), we fit the term within the brackets of the Equation (\ref{EQ:n2}) with a power-law in order to obtain the value of $p_2$ for different values of $\alpha$. The top panel of Figure (\ref{FIG:amax_p}) shows $p_2$ as a function of the radius for 3 values of $\alpha$, assuming that the original distribution has a slope $p=3.5$. The blue solid line represents a relatively low value of $\alpha$, the green dashed line of an intermediate value, and the red dash-dotted line of a high value \footnote{One refers to the magnitude of $\alpha$ compared with the Stokes number. A large, intermediate and low value of $\alpha$ in this context means $St/\alpha \gg 1, St/\alpha \sim 1$, and $ St/\alpha \ll 1$ respectively.}. One recovers the original value of $p$ for the large value of $\alpha$ as expected. For intermediate and low values of $\alpha$,  the slope $p_2$ decreases from its original value ($p$) in the inner disk and increases in the outer disk.

The bottom panel of Figure (\ref{FIG:amax_p}) shows the maximum grain size $a_{\rm max}$ needed to explain the value of $\beta$ shown in the middle panel of Figure (\ref{FIG:Disk_properties}), given  the $p_2$ curves shown in the top panel (see Appendix \ref{APP:Beta}). Note that in these cases, the maximum grain size $a_{\rm max}$ increases with $\alpha$.  
Therefore, any physical process that changes the slope $p \rightarrow p_2$, will alter the inferred value of $a_{\rm max}$.

\begin{figure}[!t]
\centering
\includegraphics[scale=0.6]{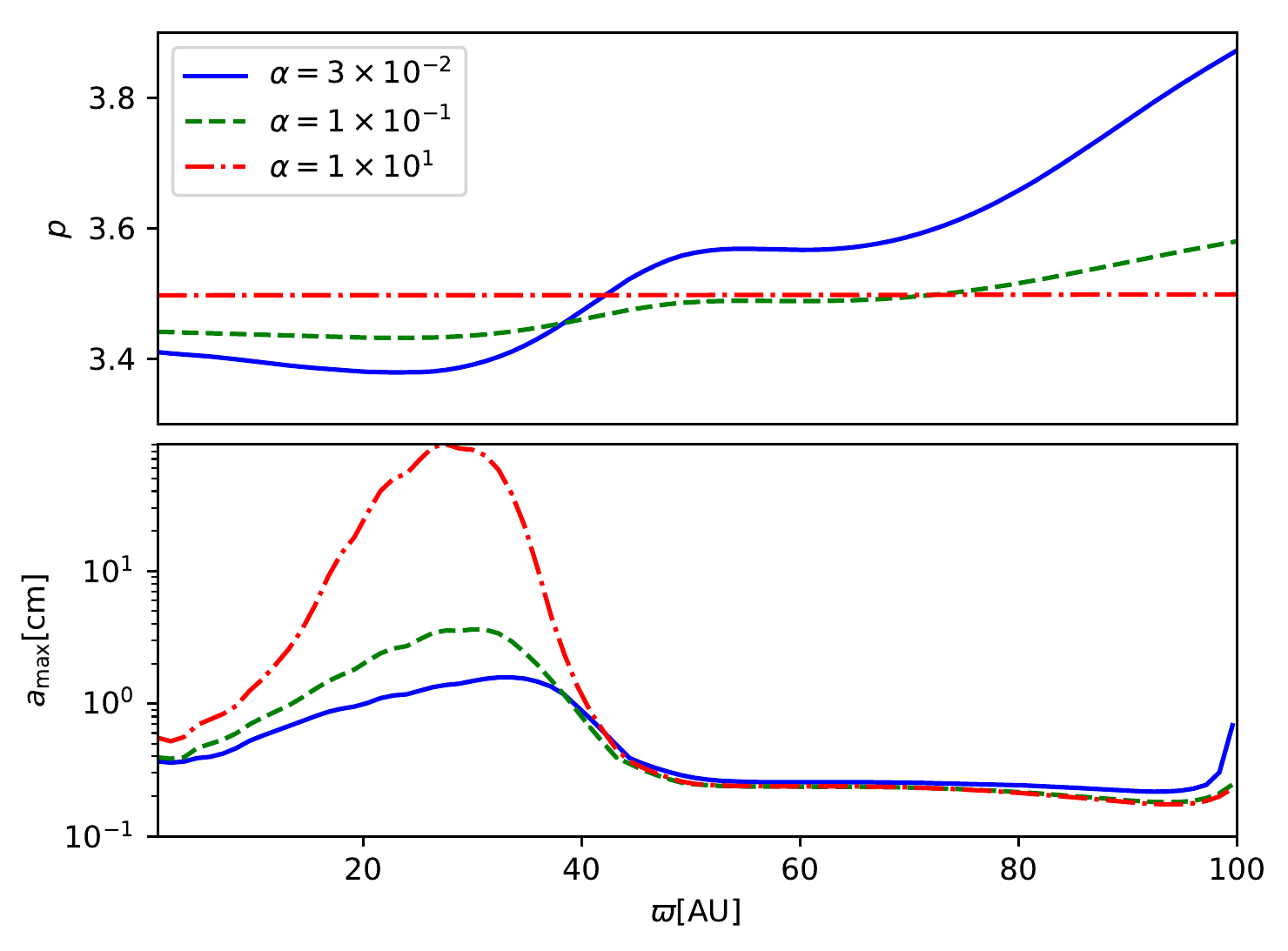}
\caption{Degeneration between $p$ and $a_{\rm max}$ to explain the profile of $\beta$ in the HD 169142 disk.}
\label{FIG:amax_p}
\end{figure}

\section{Conclusions}
\label{SEC:Conclusions}
We present an analytic model of the dust concentration on gas pressure maxima in disk rings. This model assumes steady state and only considers the dust radial redistribution, without fragmentation or coagulation.  
In addition, the model does not consider the back reaction of the dust on the gas, which is negligible if the local dust-to-gas mass ratio is smaller than 1. In the case of HD 169142, this condition is satisfied.

The inward dust migration can be stopped (or retarded) by axi-symmetric gas pressure maxima, which act as  dust traps. 
The dust grains concentrate around the gas pressure maxima and change the local dust properties (the particle size distribution, the opacity, the dust-to-gas mass ratio). 
 The dust concentration depends on the grain size, while the small grains are well coupled with the gas, the large grains tend to decouple and concentrate more around the pressure maxima due to the drag force. Diffusion prevents a strong concentration of dust grains around gas maxima. If the viscosity coefficient $\alpha$ is large, then the dust grains of all sizes tend to follow the gas distribution. If the viscosity coefficient is small, the large dust grains (those with $St/\alpha \gg 1$) strongly concentrate around the pressure maxima, while the small grains (those with $St/\alpha \ll 1$) are well mixed with the gas
(e.g., \cite{Birnstiel_2013}, \cite{Lyra_2013}, \cite{Dullemond_2018arXiv}).
This migration has important consequences in the emergent intensity and the maps of the disks. 
To model the disk emission we have considered spherical grains with a global particle size distribution  $n(a) da \propto a^{-3.5} da $, and the composition given by \cite{Pollack_1994}, which determine the values of opacity exponent $\beta$ in Figure (\ref{FIG:Beta_Map}) used to interpret the local physical conditions of the grains. With this in mind, we summarize below our main conclusions.

\begin{enumerate}
\item Our  analytical model can explain the behaviour of the dust grains trapped in a gas ring of the simulation performed by \cite{Flock_2015}, \cite{Ruge_2016}. 
It can also explain the main characteristics of the thermal dust emission of the disk around the HD 169142 star. Because in both cases the gas surface density has local maxima, 
an equilibrium between the drag force and diffusion can be reached at these maxima, allowing the accumulation of dust and preventing its migration toward the central star. 

\item This model can be easily implemented in numerical simulations that follow only the gas in protoplanetary disks to account for the dust concentration, saving computational time.

\item The redistribution of dust grains effectively increases the dust-to-gas mass ratio and the thermal dust emission within the gas pressure maxima. However, in the case of large grains ($a_{\rm max} > 1 \rm \ cm$ for millimeter observations), even though they provide most of the mass, the emergent intensity does not increase linearly with the dust-to-gas mass ratio because these grains have a low opacity at mm wavelengths. 

\item The small spectral index derived from the ALMA observations in the inner ring of HD 169142 ($10 \lesssim \varpi \lesssim 40$ AU) can be explained by the presence of large grains. The maximum dust size  in the center of the inner ring ($\sim 27$ AU) is  $a_{\rm max} > 10$ cm. This value monotonically decreases, reaching a value of $\sim 2$ mm in the outer region of the disk. 

\item The best parameters that describe the dust dynamics in the disk around HD 169142 according to the analytical model described in Section \ref{SEC:Analytical_model} are a global dust-to-gas mass ratio $\epsilon_{\rm global} = 1.05 \times 10^{-2}$ (similar to the ISM), and a viscosity coefficient $\alpha = 1.35\times 10^{-2}$. These parameters can simultaneously explain the dust emission profiles at $\lambda = 870\ \mu$m, $1.3$ mm, and $3.0$ mm.
\end{enumerate}

\textit{Acknowledgements.} 
A. S. and S. Lizano acknowledge support from PAPIIT-UNAM IN101418 and CONACyT 23863. 
M.O. acknowledges financial support from the MINECO (Spain) AYA2017-84390-C2 grant (co-funded by FEDER) and also from the State Agency for Research of the Spanish MCIU through the ``Center of Excellence Severo Ochoa" award for the Instituto de Astrofísica de Andalucía (SEV-2017-0709)”.
MF acknowledges funding from the European Research Council (ERC) under the European Union’s Horizon 2020 research and innovation programme (grant agreement No. 757957). 
We thank useful comments from an anonymous referee that helped clarified some aspects of the paper.

This paper makes use of the following ALMA data: \\ ADS/JAO.ALMA\#2013.1.00592.S, \\ ADS/JAO.ALMA\#2012.1.00799.S, \\ ADS/JAO.ALMA\#2016.1.01158.S. \\ ALMA is a partnership of ESO (representing its member states), NSF (USA) and NINS (Japan), together with NRC (Canada) and NSC and ASIAA (Taiwan) and KASI (Republic of Korea), in cooperation with the Republic of Chile. The Joint ALMA Observatory is operated by ESO, AUI/NRAO and NAOJ.
\software{CASA (v 4.7.0) \citep{McMullin_2007}}

\appendix
\section{Spectral index $\beta$ and dust opacity}
\label{APP:Beta}
The spectral index $\beta$ between $870\ \mu$m and $7$ mm is computed from the opacities obtained from the DA01 code using the \cite{Pollack_1994} abundances, in a grid varying the maximum grain size ($a_{\rm max}$) and slope $p$ of the particle size distribution $n(a) \propto a^{-p}$. The minimum grain size in all the cases is set to 0.05$\mu$m, this value is relevant only in the case when the mass of the particle size distribution is dominated by the small grains, which occurs when $p > 4$.

Left panel of Figure (\ref{FIG:Beta_Map}) shows the map of $\beta$ as a function of $a_{\rm max}$ and $p$.  Note that in the region of
small dust grains with $a_{\rm max}  \lesssim 60 \mu$m or  high slope $p \gtrsim 4$,  the value is $\beta \sim 1.8$, which is characteristic of the interstellar medium \citep{Draine_2006}. The region with $\beta < 1$ (blue) can only be explained by large dust grains $a_{\rm max}  \gtrsim 2$ mm and $p \lesssim 4$. The black dashed line is the isocontour where $\beta = 0.5$. 
The slope $\beta$ shows a local maximum for intermediate particles ($60 \mu \mathrm{m} < a_{\rm max} < 2$ mm and $p \lesssim 4$), which has been previously reported in the literature (e.g., \cite{Ricci_2010}).

Right panel of Figure (\ref{FIG:Beta_Map}) shows as an example the map of the total dust opacity $\chi$ at $\lambda = 3$ mm as a function of $a_{\rm max}$ and $p$; the units of $\chi_{\rm 3mm}$ are cm$^2$ per gram of dust. The dust opacity has a local maximum for $a_{\rm max} \sim 0.13$ cm and low $p$; this occurs because when $a_{\rm max} \gg \lambda$ the largest dust grains do not contribute to the effective cross section but they dominate the dust mass. In general, when $a_{\rm max} \gg \lambda$, the opacity decreases as $\chi_{\nu} \propto a_{\rm max}^{3-p}$ for $3<p<4$ (DA01).

Also, the dust opacity is low for large values of the slope of the particle size distribution ($p \gtrsim  4$), this occurs because for $p>4$ the mass is dominated by the small grains, however they do not have a large effective cross section at millimeter wavelengths. For small grains $\log (a_{\rm max})< 1.5$, the opacity at $\lambda = 3$ mm is also small for all the values of $p$, due to their effective cross section at millimeter wavelengths is small.

The behaviour of the dust opacity at other millimeter wavelengths is qualitatively the same, the difference is the value of $a_{\rm max}$ where the opacity is maximum. We find that  total opacity has a maximum at  $a_{\rm max} \sim \lambda /\pi$.  The maps of Figure (\ref{FIG:Beta_Map}) can be used to find the opacity at other wavelengths between $\lambda = 870 \ \mu$m and 7 mm via $\chi_{\nu} = \chi_{\rm 3 mm} (\nu/\nu_{\rm 3mm})^{\beta}$.

\begin{figure}[!t]
\centering
\includegraphics[scale=0.5]{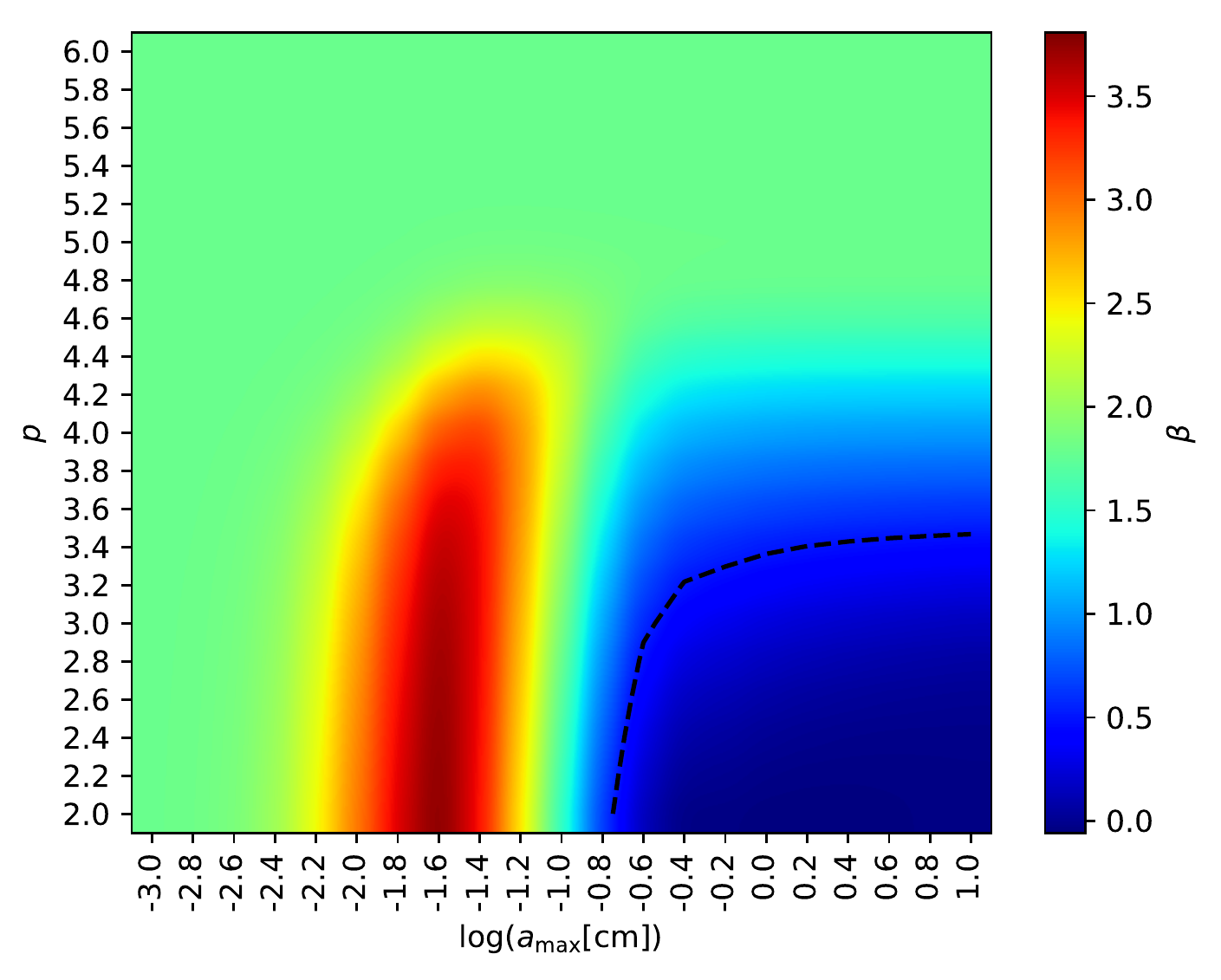}
\includegraphics[scale=0.5]{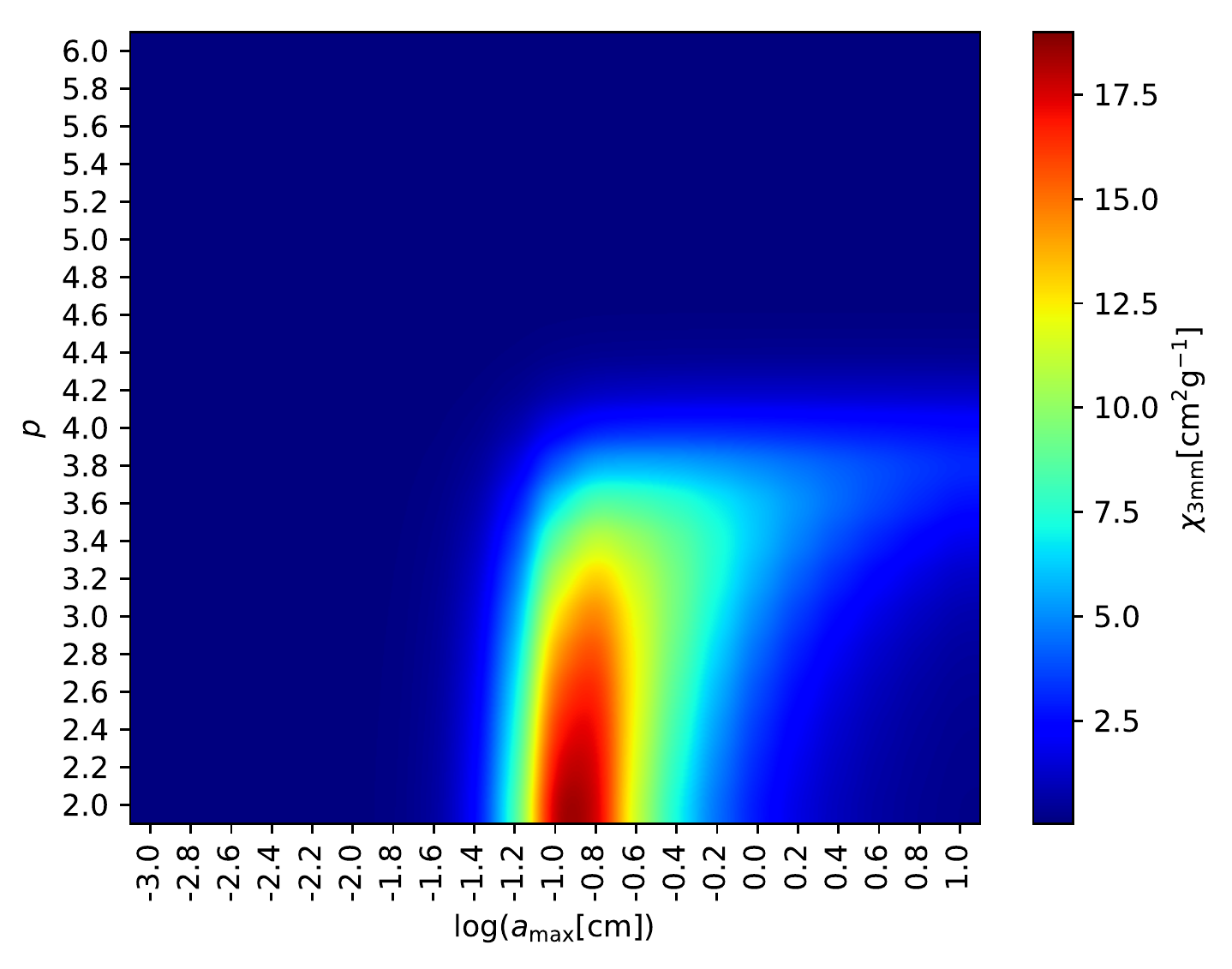}
\caption{Left panel: $\beta$ between $\lambda = 870 \mu$m and $7$ mm as a function of $a_{\rm max}$ and $p$. The black dashed line corresponds to the isocontour where $\beta = 0.5$. Right panel: Dust total opacity (absorption + scattering) at $\lambda = 3$ mm for the same space parameter. Both maps were computed using the DA01 code.}
\label{FIG:Beta_Map}
\end{figure}

\section{Gas and dust timescales}
\label{APP:Timescales}
The dynamics of the gas and dust grains is different due to the drag force. In particular,  the advection timescale of the dust grains is
\begin{equation}
t_{\rm adv,d} = \frac{L}{v_{\varpi}},
\end{equation}
where $L$ is a characteristic length scale and $v_{\varpi}$ is the magnitude of the dust velocity relative to the gas 
(Equation \ref{EQ:dust_vel}). The diffusion timescale of the gas and the dust grains are 
\begin{equation}
t_{\rm diff, g} = \frac{L^2}{D_{\rm g}}.
\end{equation}
\begin{equation}
t_{\rm diff, d} = \frac{L^2}{D_{\rm g}} (1 + St^2),
\end{equation}
where $D_{\rm g}$ is the gas diffusion coefficient and $St$ the Stokes number 
 \citep{Youding_2007}. Then, the ratio between the diffusion and advection timescales of the dust is given by
\begin{equation}
\frac{t_{\rm diff,d}}{t_{\rm adv, d}} = \frac{St}{\alpha} \left| \frac{\varpi}{\Sigma_{\rm g}} \frac{d \Sigma_{\rm g} }{d \varpi} \right|,
\end{equation}
If this ratio is much larger than 1, the dust concentrates in pressure maxima.
The ratio between the diffusion timescale of the gas and the dust advection timescale is
\begin{equation}
\frac{t_{\rm diff,g}}{t_{\rm adv, d}} = \frac{St}{\alpha} \left(\frac{1}{1+St^2} \right) \left| \frac{\varpi}{\Sigma_{\rm g}} \frac{d \Sigma_{\rm g} }{d \varpi} \right|.
\end{equation}
If this ratio is much larger than 1, as the gas evolves, the dust quickly adjusts to the gas density structure and 
concentrates in gas pressure maxima.

In general,  the Stokes number for millimeter  and centimeter dust grains ($St({\rm 1 \ mm})  \sim 0.1$, $St({\rm 1 \ cm})  \sim 1$). Also, the logarithmic density gradient in accretion disks $ \left | (\varpi/\Sigma_{\rm g})(d\Sigma_{\rm g}/d\varpi) \right|$) is in general, of the order of 1, but in pressure maxima, it is much larger than 1. For example, consider a gaussian function at a radius $\varpi_0$ and a total width $2 \sigma$, 
$\Sigma_{\rm g} \propto \exp ( - \frac{1}{2}(\varpi - \varpi_0)^2/\sigma^2)$. Then, the logarithmic  gradient is  $ \varpi |(\varpi - \varpi_0)|/\sigma^2$. 
For narrow rings such that $\epsilon = 2 \sigma / \varpi_0  <1$, the logarithmic gradient is larger than 1, except in the neighborhood of $\varpi_0$ where the dust grains are already close to the equilibrium point. Then, because the viscosity coefficient $\alpha$ is much smaller than 1,  the advection timescale of the dust grains is smaller than the diffusion timescale of both the gas and dust, allowing a fast concentration of the dust grains in the gas pressure maxima.

Figure \ref{FIG:time_scales} shows the ratio between the diffusion and advection timescales for the best fit model of the HD 169142 disk . The different lines show these ratios for dust sizes of 1 cm and 1mm. Because the Stokes number for 1 mm particles is small, the ratio of the dust diffusion to advection timescales and the ratio of the gas diffusion to the dust advection timescales are the same. 
Note that the ratios between the timescales are larger than 1 throughout the disk, with local minima at the position of the gas maxima, where the dust grains concentrate. Thus, a fast trapping scenario and the steady state assumption are feasible in the HD 169142 disk.

\begin{figure}[!t]
\centering
\includegraphics[scale=0.5]{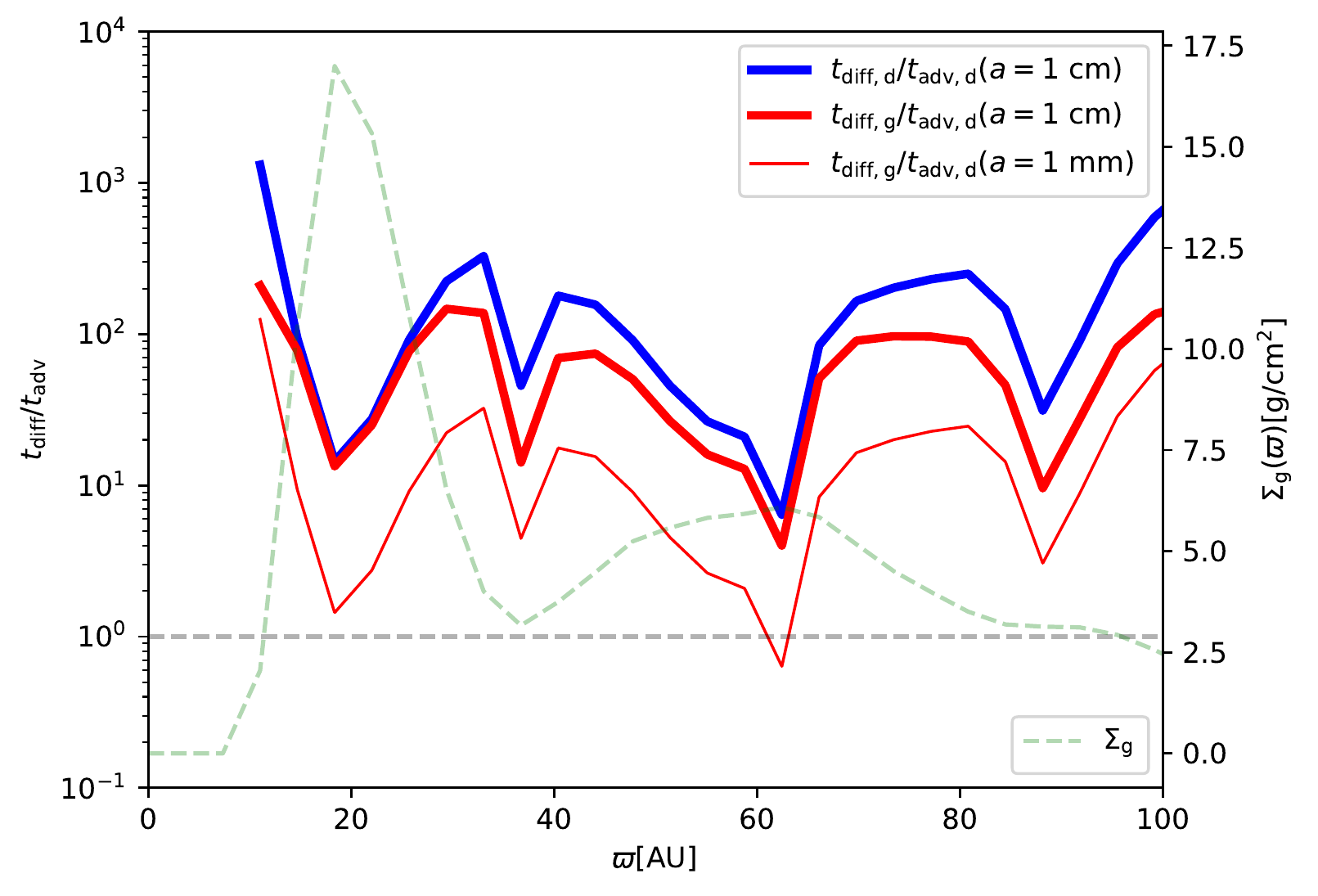}
\caption{Ratio of the diffusion and advection timescales as a function of radius for the HD 169142 disk. The thick blue line is the dust diffusion to advection timescales 
ratio for grains of 1 cm. The thick and thin red lines  are the ratio of the gas diffusion to the dust advection timescales for 1 cm and 1 mm, respectively.
The gas surface density is shown for reference in the background as a green dotted line.}
\label{FIG:time_scales}
\end{figure}
 
\section{Imaging parameters}
\label{APP:Obs_param}
Continuum images were created using CASA (v 4.7.0) \footnote{CASA, the Common Astronomy Software Applications package, is a software developed to support data processing of radio astronomical telescopes.}. Table (\ref{TAB:Obs_params}) shows the relevant imaging parameters used for each wavelength.

\begin{table}
\centering
\caption{Parameters of continuum and line Images}
\begin{tabular}{cccccccc}
& &  & &  & \multicolumn{1}{c}{Original synthesized beam size} &  \multicolumn{2}{c}{rms noise image} \\ 
\hline 
& & $\lambda$ & $\nu$ & Clean Weighting & Major x Minor; P.A. \footnote{All dust continuum images were finally convolved to the same circular beam of 0.20 arcsec.} & Original & Convolved\\ 
& Band & (mm) & (GHz) & • & (arcsec $\times$ arcsec; deg) & ($\mu$Jy/beam) & ($\mu$Jy/beam) \\ 
\hline
dust &7 & 0.87 & 330 & Briggs; robust=0.5 & 0.19$\times$0.12; 84.97 & 170 & 196 \\ 
dust & 6 & 1.3 & 230 & Natural weighting & 0.19$\times$0.13; 63.42 & 100 & 91 \\ 
dust & 3 & 3.0 & 100 & Briggs; robust=-1 & 0.17$\times$0.07;-89.22 & 25 & 23 \\ 
$^{12}$CO & 6 & 1.3 & 220 & Natural weighting & 0.16$\times$0.16, 0.0 &3650\footnote{The noise of the CO maps corresponds to the rms of the zeroth moment maps.} & -  \\
$^{13}$CO& 6 & 1.3 & 220 & Natural weighting & 0.16$\times$0.16, 0.0 & 2961 & - \\
\hline
\end{tabular}
 
\label{TAB:Obs_params}
\end{table}

\listofchanges

\end{document}